\documentclass[review]{elsarticle}
\usepackage{mathtools}
\usepackage{algorithm}
\usepackage{algorithmic}
\usepackage{lineno,hyperref}
\usepackage{amsmath}
\usepackage{amssymb}
\usepackage{amsthm}
\usepackage{algorithm}
\usepackage{algorithmic} 
\usepackage{mathtools}
\usepackage{amsthm}
\usepackage{enumerate}
\usepackage{bm}
\modulolinenumbers[5]
\usepackage{multirow}
\usepackage{booktabs}
\usepackage{graphicx}
\usepackage{longtable}
\usepackage{cleveref}
\usepackage{supertabular}
\usepackage{xltabular}
\usepackage{pdflscape}
\usepackage[table,xcdraw]{xcolor}
\usepackage[numbers]{natbib}
\usepackage{bbding}
\usepackage{caption}
\usepackage{subcaption}
\usepackage{color}

\journal{Journal of The Franklin Institute}

\begin{document}

\begin{frontmatter}

\title{Towards Efficient Communications in Federated Learning: A Contemporary Survey}

\author[mymainaddress]{Zihao Zhao}
\ead{zhao-zh21@mails.tsinghua.edu.cn}

\author[mymainaddress]{Yuzhu Mao}
\ead{myz20@mails.tsinghua.edu.cn}

\author[mythirdaddress]{Yang Liu}
\ead{liuy03@air.tsinghua.edu.cn}

\author[myfourthaddress,myfifthaddress]{Linqi Song}
\ead{linqi.song@cityu.edu.hk}

\author[mysixaddress]{Ye Ouyang}
\ead{ye.ouyang@asiainfo.com}

\author[mymainaddress,mysecondaryaddress]{Xinlei Chen}
\ead{chen.xinlei@sz.tsinghua.edu.cn}

\author[mymainaddress,mysecondaryaddress]{Wenbo Ding\corref{mycorrespondingauthor}}
\cortext[mycorrespondingauthor]{Corresponding author}
\ead{ding.wenbo@sz.tsinghua.edu.cn}


\address[mymainaddress]{Tsinghua-Berkeley Shenzhen Institute, Tsinghua Shenzhen International Graduate School, Tsinghua University, Shenzhen, Guangdong, China.}
\address[mythirdaddress]{the Institute for AI Industry Research (AIR), Tsinghua University, Beijing, China.}
\address[myfourthaddress]{City University of Hong Kong, Hong Kong, China.}
\address[myfifthaddress]{City University of Hong Kong Shenzhen Research Institute, Shenzhen, Guangdong, China.}
\address[mysixaddress]{AsiaInfo Technologies, Beijing, China.}
\address[mysecondaryaddress]{RISC-V International Open Source Laboratory, Shenzhen, Guangdong, China.}

\begin{abstract}
In the traditional distributed machine learning scenario, the user's private data is transmitted between {{clients}} and a central server, which results in {{significant}} potential privacy risks. In order to balance the issues of data privacy and joint training of models, federated learning (FL) is proposed as a {{particular}} distributed machine learning {{procedure}} with privacy protection mechanisms, which can achieve multi-party collaborative computing without revealing the original data. However, in practice, FL faces {{a variety of}} challenging communication problems. {{This review seeks to elucidate the relationship between these communication issues by methodically assessing the development of FL communication research from three perspectives: \textit{communication efficiency}, \textit{communication environment}, and \textit{communication resource allocation}.}}
Firstly, we sort out the current challenges existing in the communications of FL. {{Second, we have collated FL communications-related papers and described the overall development trend of the field based on their logical relationship. Ultimately, we discuss the future directions of research for communications in FL.}}
\end{abstract}


\end{frontmatter}

\section{Introduction}
With the advances in deep learning (DL) models, recent years have witnessed a dawn of a new era of artificial intelligence. DL is now utilized in a variety of industries, including autonomous driving \cite{grigorescu2020survey, al2017deep, muhammad2020deep} and intelligent healthcare \cite{miotto2018deep, sahoo2019deepreco, philip2021deep}. However, as the size of datasets and the complexity of the newly proposed neural networks increase, training DL models becomes significantly difficult. Consequently, several methods are offered so as to accelerate the training process of DL. For one thing, maximizing the use of hardware processing resources under appropriate software control is an excellent way to shorten training time, such as data parallelism \cite{chetlur2014cudnn, sze2017efficient}. For another, distributed machine learning is devised to address this issue by separating the large-scale learning process on one workstation into several small learning processes on a number of distributed workstations, which has been the most frequently adopted method in recent years. Notwithstanding, training data is usually fragmented and shared with different clients in most distributed machine learning procedures, whereas some data cannot be aggregated into a single central server since they are privacy-sensitive in nature. For instance, user behavior data in some online shopping websites may directly contain sensitive information, {{such as}} personal age, race, address; or it may indirectly carry implicit sensitive information, such as personal web browsing records and user political inclinations implied by content preferences. With the promulgation of privacy and data protection laws and regulations such as the General Data Protection Regulation (GDPR) and the improvement of people's awareness of privacy protection, more and more attention has been paid to the privacy and security of user data. 

Thus, \textbf{federated learning (FL)} \cite{mcmahan2017communication} is developed as a data privacy-aware distributed machine learning framework. Specifically, the client utilizes its own private data to train a local model and transmit it to the server side. {{Subsequently, the server aggregates these parameters to compute the global parameter}} and sends it back to all clients. Through the multiple rounds of learning and communication described above, FL eliminates the need to collect all private local data on a single central server, overcoming privacy and communication challenges in machine learning tasks, as the data are retained locally throughout the training process. Since the private property of FL, it is widely used in our daily life, such as mobile keyboard prediction \cite{hard2018federated}, financial fraud detection \cite{yang2019ffd}, and precision medicine \cite{rieke2020future}.

Despite the benefits that FL brings to us, it also faces several challenges. First of all, since the model information is high-frequently exchanged between clients and the server, the process is highly restricted by the communication conditions in FL. Therefore, the communication overhead is the main bottleneck of FL. Second, client drift is also a huge issue in FL. As an example, the clients suffer from: 
1) \emph{statistical (data) heterogeneity:} the data of each client may be not independent and identically distributed (non-{{iid}}); 
2) \emph{model heterogeneity:} the model structure of each client may be various.
3) \emph{resource heterogeneity:} the computation, storage, and communication resources of clients may vary from one to another.
{{These heterogeneities make the entire Fl system challenging to train. For instance, the statistical heterogeneity may limit the model convergence rate; the model heterogeneity may prevent the low bandwidth clients from receiving the cumbersome global model, resulting in the straggler issue; and the resource heterogeneity may also cause the straggler and dropout problems.}}
Moreover, one easily overlooked but equally important challenge is the privacy issue in FL. Most people think that conveying the model parameters or model gradients instead of privacy data will not cause privacy leakage. Nonetheless, it has been demonstrated that this view is incorrect \cite{Zhao2022DeepLF}, and not transmitting privacy-sensitive local data still leaves security gaps. {{Therefore, a consummate trusted-FL system needs privacy-preserving techniques to prevent the leakage of local data. However, most proposed security strategies are very time-consuming \cite{li2021survey}, which also demonstrates the necessity of efficient protection approaches.}}

In this survey, we mainly discuss the first challenge, i.e., communications in FL. {{Some previous reviews have identified the main problems in FL communication from different viewpoints and classified existing research correspondingly. \citet{shahid2021communication} considered communication costs and provided an overview of related current methods such as client selection, local updating, and compression schemes. Moreover, \citet{xu2020compressed} focused on compressed communication and introduced four compressors (quantization, sparsification, hybrid, and low-rank) in detail. Considering the training workflow of synchronized federated learning, \citet{jiang2022towards} discussed methods that aim to improve the training efficiency during different phases (client selection, configuration, and reporting) respectively. In addition to tackling the efficiency issue, \citet{yang2021federated} explored FL's applications in wireless communication and proposed to address some key open problems in wireless communication by FL methods, including communication delay, energy, reliability, and massive connectivity. 

This work aims to provide a comprehensive description of communication problems in FL systems, and summarized the state-of-the-art research in all aspects involved so far. }}Specifically, we will cover the following aspects of the communication process: a) communication efficiency, b) communication environment, and c) communication resource allocation. The main contributions of this paper are trifold:
\begin{itemize}
    \item We present a taxonomy of recent FL communication approaches and summarize the FL communication system framework with listed specific techniques in each field. 
    \item We provide a comprehensive summary of recent communication algorithms in a table and sort them out in terms of \textbf{method}, \textbf{communication}, and \textbf{evaluation} objects.
    \item We propose some potential future research directions in the field of FL communications.
\end{itemize}

\section{Problem Statement and Challenges}
\subsection{Federated Learning}
Assume an FL system contains $N$ clients and all the training data and labels constitute an input space $\{\mathcal{X}_1,\cdots,\mathcal{X}_N\}$ and a target space $\{\mathcal{Y}_1,\cdots,\mathcal{Y}_N\}$. The $i^{th}$ device in this FL system has its own local input space $\mathcal{X}_i \in \{\mathcal{X}_1,\cdots,\mathcal{X}_N\}$ and target space $\mathcal{Y}_i \in \{\mathcal{Y}_1,\cdots,\mathcal{Y}_N\}$, and will sample $m_i$ instances with $n_i$ features to build a local training dataset $\mathcal{D}_i = \{(\boldsymbol{x}^{(1)}, y^{(1)}), \cdots, (\boldsymbol{x}^{(m_i)}, y^{(m_i)})\}$ sampled from the local distribution $\mathbb{P}_i (\mathcal{X}_i, \mathcal{Y}_i)$, where $\boldsymbol{x}^{(i)} \in \mathbb{R}^{n_i}$ and $y^{(i)} \in \mathbb{R}$. 
In traditional distributed machine learning, these training datasets are collected in a central server, while all private data are stored on the {{client's}} own devices {{in FL}}.

\begin{figure}[!htbp]
    \centering
    \includegraphics[width=\linewidth]{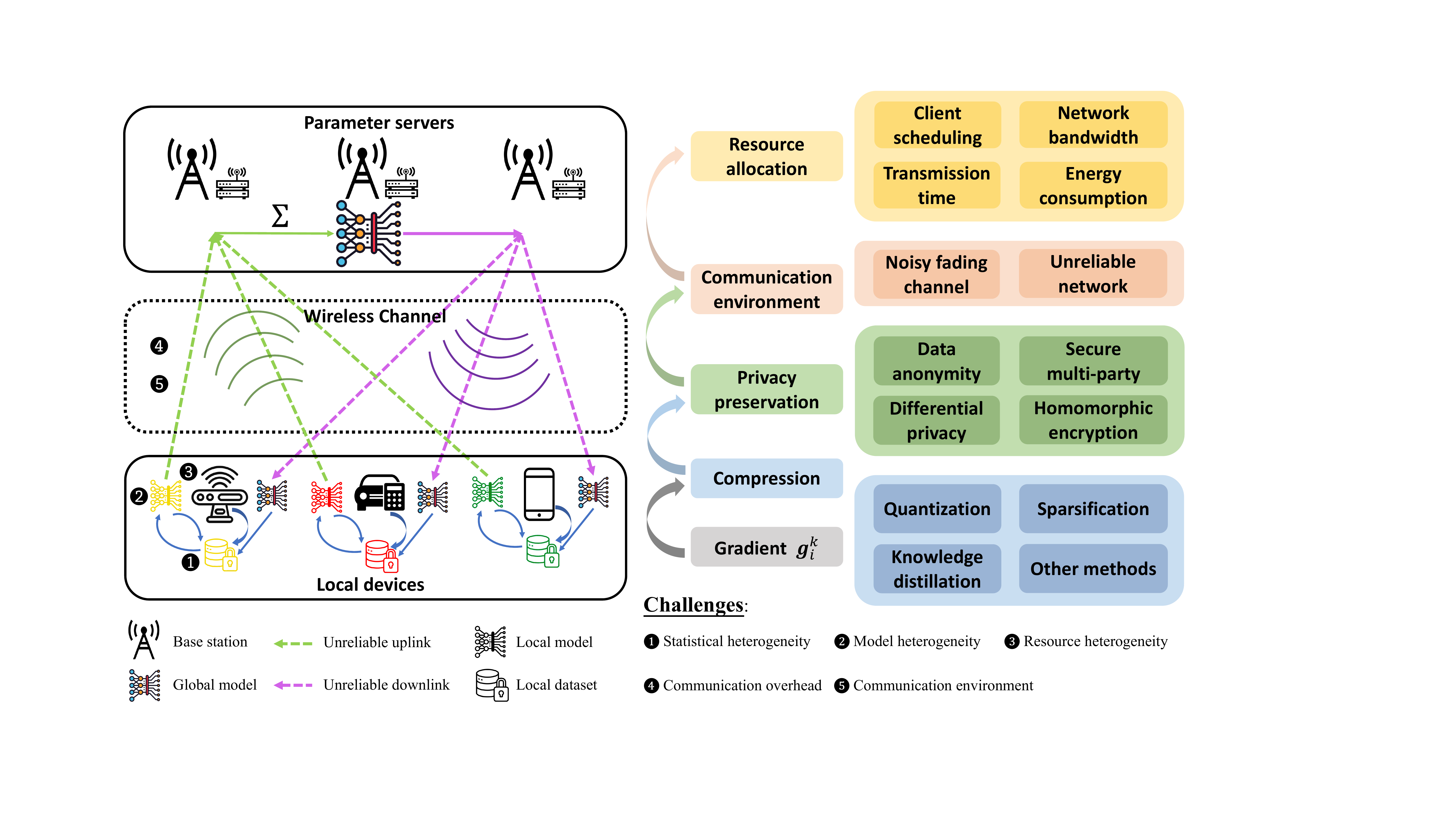}
    \caption{{{Typical FL communication framework. The left part illustrates the training procedure of FL and proposes five dominating challenges of FL communication. The right shows the related works to tackle the aforementioned issues. After each client calculates its gradient $g_i^k$ at $k$-th global epoch, a compression method could be selected to efficiently train a global model. Finishing compression, clients could apply some privacy preservation algorithms to protect their gradient information further. Since the communication environment may be imperfect and the resource of each client may be imbalanced, the server could choose dynamical allocation strategies to mitigate their severe influence to FL system convergence.}}}
    \label{framework}
\end{figure}

In FL, a group of clients train their local model $W_i$ on their local private dataset $\mathcal{D}_i$ and then transmit the training results (e.g., model parameters or gradients) to the central server. Subsequently, the server will aggregate the received results to update the global parameter or gradient and send it back to the corresponding clients, in order to facilitate their local model updates. {{The whole procedure of FL is illustrated in \Cref{framework}}} and the optimization problem of this FL system could be formulated as follows:
\begin{equation}
    \min _{\boldsymbol{W}} f(\boldsymbol{W})= \sum_{i=1}^{N} p_i f_{i}(\boldsymbol{W}), f_{i}(\boldsymbol{W})=\mathbb{E}_{(\boldsymbol{x}^{(\zeta_{i})}, \boldsymbol{y}^{(\zeta_{i})}) \sim \mathcal{D}_i}\left[\mathcal{L}\left(\mathcal{F}_{i}\left(\boldsymbol{x}^{(\zeta_{i})}; \boldsymbol{W}\right), \boldsymbol{y}^{(\zeta_{i})}\right)\right],
\end{equation}
{{where $f$ denotes the global empirical risk, $p_i$ denotes the aggregated weight of client $i$, $f_i$ denotes the local empirical risk, $\boldsymbol{W}$ denotes the model parameter}} which optimizes the above objective function, $p_i$ is the aggregation weight of each client (usually $1 / N$), {{$\mathcal{L}$ and $\mathcal{F}_i$ are the loss function and the neural network function of $i^{th}$ client, respectively, and $(\boldsymbol{x}^{(\zeta_{i})}, \boldsymbol{y}^{(\zeta_{i})})$ denotes the mini-batch samples of local dataset $D_i$.}} In the most common FedAvg algorithm \cite{mcmahan2017communication}, the optimization objective $\boldsymbol{W}$ represents the global {{parameter}} $\boldsymbol{W_g}$ aggregated by the model {{parameter $\{\boldsymbol{W}_i\}_{i=1}^{N}$}} of all clients. The whole FL training process (transmitting model parameter) is shown in \Cref{FedAvg}. Note that the optimal solution of the global empirical risk $f$ and the local empirical risk $f_i$ could be different. {{We define $f^{*} = \min _{\boldsymbol{W_g}} f(\boldsymbol{W_g})=f(\boldsymbol{W_g^{*}})$ and $f_i^{*} = \min _{\boldsymbol{W}_i} f_i(\boldsymbol{W}_i)=f_i(\boldsymbol{W_i^{*}})=\mathbb{E}_{(\boldsymbol{x}^{(\zeta_{i})}, \boldsymbol{y}^{(\zeta_{i})}) \sim \mathcal{D}_i}\left[[\mathcal{L}\left(\mathcal{F}_{i}\left(\boldsymbol{x}^{(\zeta_{i})}; \boldsymbol{W}_i^*\right), \boldsymbol{y}^{(\zeta_{i})}\right)\right]$.}}
\subsection{Recent communication challenges in FL}

\begin{itemize}
    \item \textbf{Statistical (data) Heterogeneity.}
    Most traditional deep learning techniques, such as face recognition \cite{zhang2020fedpd, arandjelovic2005face} and object detection \cite{yu2019federated}, assume that the training data are independent and identically distributed ({{iid}}). However, in practice, most of the training data are {{non-iid}}, and they will significantly influence the convergence rate of the entire FL process, potentially {{exacerbating}} communication overhead. In the FL setting, non-iidness means that the training data distributions of clients are different:
    \begin{equation}
        \mathbb{P}_i (\mathcal{X}_i, \mathcal{Y}_i) \neq \mathbb{P}_j (\mathcal{X}_j, \mathcal{Y}_j),
    \end{equation}
    for different $i^{th}$ and $j^{th}$ client. Moreover, some research \cite{chen2021fedsa} also considers that the data are {{non-iid}} if the expectation of local gradients and global gradients are different:
    \begin{equation}
        \mathbb{E}\left[\left\|\boldsymbol{g}_{i}^{t}-\bar{\boldsymbol{g}}^{t}\right\|\right] \neq 0,
    \end{equation}
    where $g_i^t$ {{denoets the uploaded gradient}} of the $i^{th}$ client and $\bar{g}^{t}$ {{denotes the average global gradient}}.
    
    \item \textbf{Model Heterogeneity.}
    Since the resources of participants in FL vary widely, the size of the model they are able to train can also be different. Therefore, in each epoch, the {{uploaded}} model structure may be different:
    \begin{equation}
        \operatorname{shape}(\boldsymbol{W}_i) \neq \operatorname{shape}(\boldsymbol{W}_j),
    \end{equation}
    where the $\operatorname{shape}(\cdot)$ operator outputs the shape of each input model {{$\boldsymbol{W}_i$.}}
    
   \item \textbf{Resource Heterogeneity.}
    Due to the variety of different clients and communication environments, FL will be challenged in different ways. For example, the transmission channel may be noisy and fading, and the bandwidth $B$ of the channel may be limited. Furthermore, the energy consumption $E$ and time latency $T$ of each participant may be constrained. Specifically, the energy consumption contains the energy of transmitting data $E^U$, receiving data $E^R$, and local computation and training $E^C$. Moreover, the time latency may also include uploading time $T^U$, receiving time $T^R$, and computing and training time $T^C$. Thus, the optimization problem can be formulated as follows:
    \begin{equation}
    \begin{array}{ll}
    \min  & f(\boldsymbol{W_g}) \\
    \text { s.t. } & B \leq \Tilde{B} \\
    & E^U + E^R + E^C \leq \Tilde{E} \\
    & T^U + T^R + T^C \leq \Tilde{T},
    \end{array}
    \end{equation}
    where the $\Tilde{B}, \Tilde{E}, \Tilde{T}$ denote the bandwidth, energy and time budget, respectively.
    
    \item \textbf{Communication Overhead.}
    As the size of recently proposed neural networks increases, the communication process in FL becomes slower and slower. Thus, many communication-efficient FL algorithms are proposed to tackle this issue, and their pattern could be summarized as follows. Suppose client $i$ has a dense {{model parameter $\boldsymbol{W}_i$}} to transmit, and $Q$ denotes the compression operator to compress {{$\boldsymbol{W}_i$ to a sparse one $Q(\boldsymbol{W}_i)$}} and thus reduce communication cost. The optimization problem {{for a compression strategy $Q$ and model parameters $\boldsymbol{W}_i$}} could be formulated as:{{
    \begin{equation}
        \min _{Q, \boldsymbol{W}_i} f^Q_i(\boldsymbol{W}_i) + \lambda \operatorname{Bit}(Q(\boldsymbol{W})) + \mu \|\boldsymbol{W}_i-Q(\boldsymbol{W}_i)\|_{2}^{2},
    \end{equation}}}
    where $f^Q_i$ denotes the $i$-th client's loss function of the compressed network, {{$\operatorname{Bit}(\cdot)$ denotes the summation of transmitted bits, and $\lambda, \mu > 0$ represent tuning hyper-parameters.}} The above problem seeks to optimize the model performance while subject to the constraint of a compression error regularizer. To solve this multi-goal optimization problem, a common method is the alternating direction method of multipliers (ADMM) \cite{boyd2011distributed} instead of stochastic gradient descent (SGD) \cite{bottou2010large}.

    \item \textbf{Communication Environment.} 
    In a realistic communication environment, the communication channel is not perfect, and it may be noisy and fading, which slows down the convergence of model aggregation and reduces the performance of the global model. Assuming the transmitted global model is $\boldsymbol{W_g}$ and the received global model of clients is $\boldsymbol{W_g}^{\prime}$, then the channel condition could be denoted as:
    \begin{equation}
        \boldsymbol{W_g}^{\prime} = h \boldsymbol{W_g} + z,
    \end{equation}
    where $h$ denotes the coefficients of the fading channel and $z$ represents the additive channel noise.
        
\end{itemize}

\begin{figure}
    \centering
    \includegraphics[width=1\linewidth]{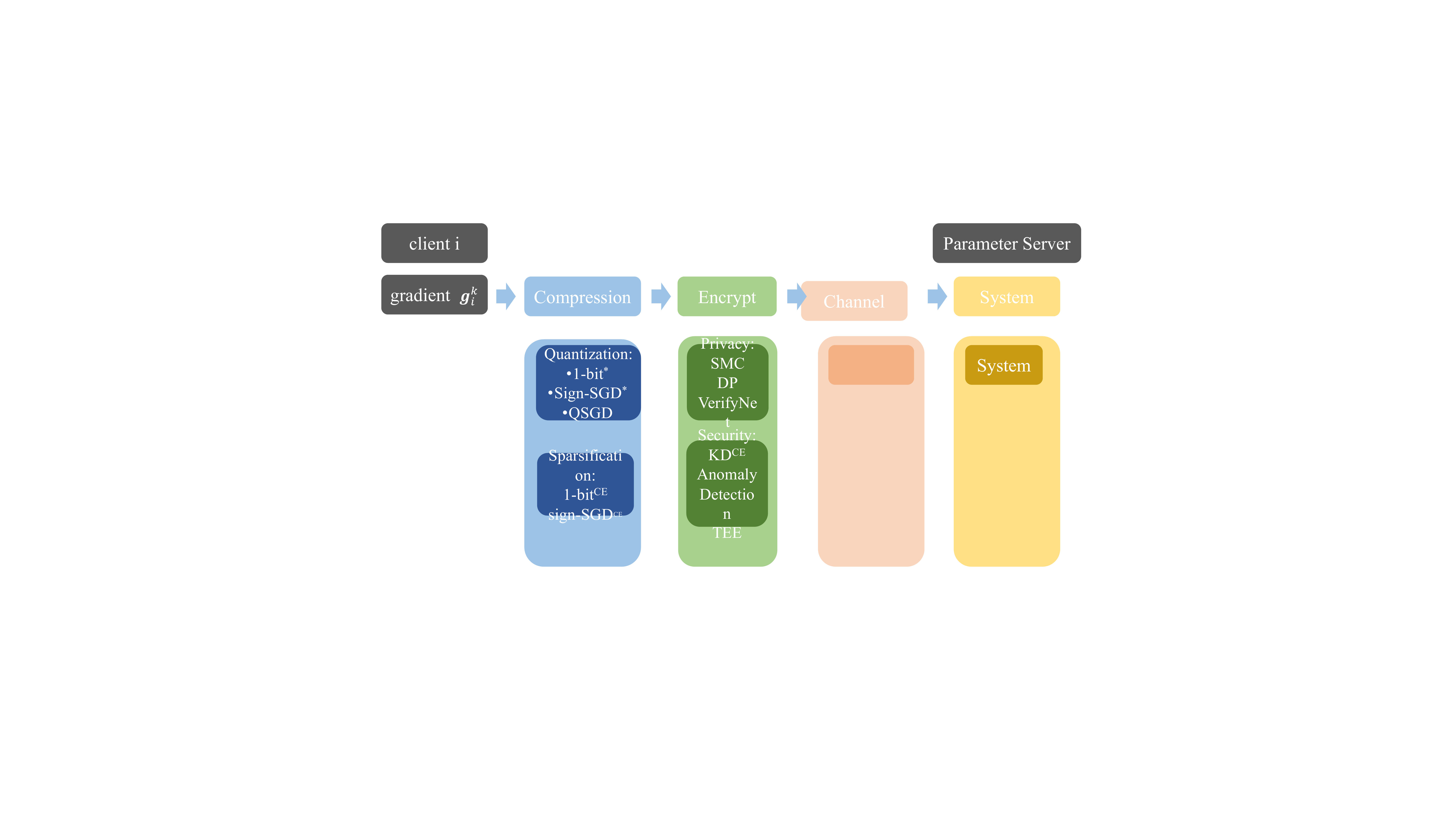}
    \caption{{{The primary structure of this review's body part based on the categorization of communications in FL.}}}
    \label{chap_comm_eff}
\end{figure}

\begin{algorithm}[tb]
  \caption{An example of FL training procedure (sending{{ model parameters}})}
  \label{FedAvg}
\begin{algorithmic}
  \STATE {\bfseries Input:} The entire $N$ clients are indexed by $i \in \{1, 2, \cdots, N\}$; $D_{i}=\left\{\left(\boldsymbol{x}^{(i)}, y^{(i)}\right)\right\}_{i=1}^{n_{i}}$ is the local dataset of client $i$; $T_g$ and $T_{loc}$ is the number of global epochs and local epochs, respectively, and $\alpha$ is the learning rate.
     \STATE {\bfseries Server executes:} 
        \begin{ALC@g}
        \STATE Initialize $\boldsymbol{W}_g^0$ 
         \qquad \FOR{each round $t = 1, 2, \cdots, T_g$}
        
            \FOR{each client $i$ {\bfseries in parallel}}
                \STATE$\boldsymbol{W}_i^{t+1} \leftarrow ClientUpdate(i, \boldsymbol{W}_g^t)$
            \ENDFOR
            \STATE $\boldsymbol{W}_g^{t+1} \leftarrow \frac{1}{N}\sum_{i=1}^{N} \boldsymbol{W}_i^{t+1}$
        \ENDFOR
         \end{ALC@g}
     \STATE {\textit{ClientUpdate}($i, \boldsymbol{W}_g^t$):}

    \FOR{each local epoch from $1$ to $T_{loc}$}
            \STATE $\boldsymbol{W}_i^{t+1} \leftarrow \boldsymbol{W}_g^t-\alpha \nabla f_i(\boldsymbol{W}_g^t, D_i)$
    \ENDFOR
    \STATE Return $\boldsymbol{W}_i^{t+1}$ to the server

\end{algorithmic}
\end{algorithm}
                                         
\section{Communication Efficiency}
In this section, we summarize three types of most commonly used communication-efficient FL methods into three parts, which are {{quantization-based, sparsification-based, and distillation-based strategies}}. {{Specifically, for quantization-based methods, we focus on the diverse designs of quantization operators, which give various ways of transforming a floating-point of 64/32 bits to a lower precision and thus determine the intrinsic property as well as the theoretical foundation for different FL frameworks featuring quantization. For sparsification-based methods, we classify existing works from the viewpoint of framework design, which provides different solutions to determine nonzero components.  Moreover, given the special challenge of privacy protection in FL systems, the distillation-based strategies are classified into data-additional and data-free categories based on distinct levels of potential privacy leakage. Note that most of these communication methods are conducted to the transmitted gradients rather than model parameters.}} Additionally, there is also {{an extra section}} for other minor methods that are not frequently used. Finally, we elucidate the comparison of the cited methodologies in \Cref{Table1}.

\subsection{Quantization}
Quantization \cite{gray1998quantization} is a technique that decreases the model size by representing the bit width from a floating-point of 32 bits to a lower precision meanwhile retaining the model performance. Particularly, in the FL scenario, most quantization methods are proposed to compress the continuous model gradient value of each client into a discrete set after the local training process so as to reduce the representing bit. 

\textbf{Stochastic Quantization.} 
The stochastic quantization (SQ) is introduced in \citet{alistarh2017qsgd}, which uses a gradient quantization method called QSGD to improve the communication transmission problem in parallel SGD computing, and focuses on solving the trade-off between the transmission channel bandwidth and convergence time. Specifically, the quantizer $\operatorname{Q_{SQ}}$ in \cite{alistarh2017qsgd} is defined as:
\begin{equation}
\operatorname{Q_{SD}}\left(g_{i}\right)=\|\boldsymbol{g}\|_{2} \cdot \operatorname{sign}\left(g_{i}\right) \cdot\left\{\begin{array}{l}
\ell / s, \text { with probability } 1-\frac{g_{i}}{\|\boldsymbol{g}\|_{2}} s+\ell \\
(\ell+1) / s, \ o . w .
\end{array}\right. ,
\end{equation}
where $s \ge 1$, $0 \le \ell < s$ are two tuning hyperparameters and $\|\cdot\|_2$ is the $l^2$-norm.
By doing so, it preserves the statistical properties of the primary vector and introduces minimal variance. Then some QSGD-based variation algorithms are proposed. \citet{reisizadeh2020fedpaq} introduce a framework called FedPAQ, which quantizes model updates by QSGD under a restricted partial {{client}} participation circumstance and reduces communication rounds by setting the synchronization of each client with the parameter server periodically. As an extension of FedPAQ, \cite{haddadpour2021federated} utilize the historical information of global models and theoretically analyze the proposed FedCOM under both homogeneous and heterogeneous local data.
Furthermore, \citet{das2020faster} consider both heterogeneous local data and various noises of local stochastic gradients and propose FedGLOMO to reduce the variance of local updates by global aggregation {{with momentum}}. \citet{dai2019hyper} present hyper-sphere quantization (HSQ) to build a global cookbook and quantize local updates based on this cookbook and SQ quantizer to reduce the communication cost {{further}}. 

However, in the paper mentioned above, the quantization level could not dynamically change during the entire FL training process. To this end, \citet{jhunjhunwala2021adaptive} propose an adaptive quantization algorithm called AdaQuantFL, considering the trade-off between error and communication bits and allowing clients to automatically adjust the quantization level $s$ of QSGD during the entire FL training process. Moreover, \citet{amiri2020federated} propose a lossy FL (LFL) approach to quantize both the global model {{parameters}} and the client model {{parameters}} to reduce the communication cost {{further}}, while most previous work assumes that the broadcast of the global model is perfect.

\textbf{Rotation-based Quantization.} Although SQ is an efficient and convenient quantization strategy to assign each vector coordinate to a finite set of possibilities, recent research shows that SQ is sensitive to the vector distribution and the gap between the largest and smallest entries in the vector. Therefore, lots of studies concentrate on solving the Distributed Mean Estimation (DEM) problem. Specifically, \citet{suresh2017distributed} devise a biased and deterministic quantization framework by applying a structured random rotation before quantization. Following this method, \citet{vargaftik2021drive} introduce a biased and unbiased compression technique, DRIVE, which could quantize the original vector into a 1-bit quantization level by random rotation in DEM. Although the previous work mostly defines the quantization through a set of discrete quantization points, \citet{vargaftik2022eden} build the theory on an interval-wise quantization and proposes EDEN, in which each coordinate is quantized to its interval's center of mass rather than the nearest quantization point. Therefrom, it decreases the entropy of the quantized vector and obtains a better estimation given a communication budget. Nonetheless, previous work requires that each client has an independent rotation matrix from other clients, which asymptotically increases the decoding time as the server must invert the rotation for each independent vector in every iteration.
\citet{basat2022quick} propose a strategy named QUIC-FL to speed up the aggregation, which enables all clients to utilize the same rotation matrix, resulting in only a single inverse rotation on the {{server side.}}

\textbf{Lattice-based Quantization.}
\citet{zamir1992universal} first introduce the lattice quantization with dithers to optimize every point to its closest lattice, which is much simpler and more efficient than optimal vector quantizers. \citet{shlezinger2020federated} firstly bring the lattice quantization into the FL framework. Let $\mathcal{L}$ be the lattice set, then the lattice quantizer $\operatorname{Q_{\mathcal{L}}}$ is defined by:
\begin{equation}
\operatorname{Q_{\mathcal{L}}}(\boldsymbol{g})=\boldsymbol{l}_{\boldsymbol{x}} \text { if }\left\|\boldsymbol{g}-\boldsymbol{l}_{\boldsymbol{g}}\right\| \leq\|\boldsymbol{g}-\boldsymbol{l}\| \text { for every } \boldsymbol{l} \in \mathcal{L} .
\end{equation}
Subsequently, \citet{shlezinger2020uveqfed} take a more realistic and comprehensive situation into consideration and offer a universal vector quantization in FL called UVeQFed. Specifically, they consider the rate-constrained channels and implement the subtractive dithered lattice quantization to tackle the throughput-limited uplink problem and finally induce only a minimum distortion. Furthermore, except for channels with restricted rates, \citet{chen2021communication} further incorporate lattice quantization by selecting a subset of clients according to their probability of connection to the server and allowing the server to allocate bandwidth to minimize transmission delay and optimize the usage of wireless channels.

\textbf{1-bit(Sign) Quantization.}
Previous quantization work usually compresses model updates or model gradients into a low-precise bit width, such as 8 bits or 4 bits. Despite this, \citet{bernstein2018signsgd} introduce a 1-bit quantization method named signSGD that represents the original gradient by its sign meanwhile maintaining the testing accuracy and convergence rate of the model, and its quantizer $\operatorname{Q_{sign}}$ is defined as:
\begin{equation}
\operatorname{Q_{sign}}({\boldsymbol{g}})=\frac{{\boldsymbol{g}}}{\sqrt{{\boldsymbol{g}}^{2}}}.
\end{equation}
\citet{jin2020stochastic} propose sto-signSGD that the quantization result of the gradient is a random variable corresponding to its sign rather than always fix, and then combine it with the majority vote rule to ensure the robustness of sto-signSGD under the client data heterogeneity situation in FL. 
\citet{zhu2020one} employ the signSGD for Federated edge learning (FEEL) and design a sophisticated FEEL framework termed OBDA based on over-the-air majority-vote, which incorporate 1-bit gradient quantization and QAM modulation to archive communication efficiency.

\textbf{Quantized Compressed Sensing (QCS).}
By taking advantage of the sparsity of the signal, compressed sensing (CS) \cite{donoho2006compressed} combines the sampling and compression stages of traditional signal processing methods into one step, thereby greatly reducing the number of samples required for accurate signal recovery.
Specifically, CS first utilizes the matrix transformation to obtain a sparse representation of the primary dense signal and then reconstructs it from the observed data information by solving an optimization problem. \citet{abdi2020quantized} first combine CS with quantization in a distributed deep learning scenario to obtain arbitrarily large compression gains. In particular, this paper introduced the QCS framework to solve the increasing gradient variance and the decrease in the convergence rate induced by quantization. 
\citet{li2021communication} and \citet{fan2021communication} develop a 1-bit QCS by utilizing binary iterative hard thresholding (BIHT) to reconstruct model updates/gradients rather than simply minimizing the mean squared error (MSE) \cite{abdi2020quantized}. However, since these two algorithms only use 1-bit quantization, they have a large quantization error due to this limitation. Therefore, \citet{oh2021quantized} turn to study the multi-level scalar quantization method, which develops a framework called Q-EM-GAMP based on the expectation-maximization (EM) algorithm to serve as the reconstruction strategy to reduce the reconstruction error {{significantly.}}

\textbf{Others.} With the exception of the quantization algorithms mentioned above, there is still a variety of quantization 
For instance, without directly using the sign of the model gradient, \citet{DBLP:journals/corr/abs-2012-08241} establish a non-uniform cosine-based quantization called CosSGD, which quantizes the angle vector with respect to the model gradient by incorporating cosine functions and the range of its minimal and maximal value. In addition, \citet{malekijoo2021fedzip} design FedZip which uses the quantization with k-means clustering as well as combining Top-z sparsification and Huffman encoding to maximize the compression rate. What's more, in order to solve the {{non-iid}} problem in FL, \citet{philippenko2020bidirectional} present Artemis which first compresses the model gradient by its memory term and then utilizes s-quantization to compress the difference. They find that using the memory mechanism could improve the convergence performance under the {{non-iid}} training data scenarios.

{{\citet{chen2021dynamic} consider the heterogeneous FL scenario that local clients hold various quantization levels (precision) and propose FedHQ to assign different aggregation weights to clients by optimizing the convergence upper bound. 
\citet{cui2021slashing} design the MUCSC algorithm to compress the uploads by soft clustering and provide some theoretical properties of MUCSC. In addition, they also introduce a boosted MUCSC to tackle the situation with rather scarce network resource.}}

\subsection{Sparsification}
Sparsification techniques transform a full gradient to a sparse one with a subset of important elements and set other insignificant coordinates to zero. In a federated learning regime, top-$k$ sparsification and rand-$k$ sparsification are two commonly adopted methods \cite{eghlidi2020sparse}. In top-$k$ sparsification, the subset of remained elements consists of $k$ percent of {{the sparsification target}} with the greatest absolute values, while the remained values are selected randomly in rand-$k$ sparsification. Although rand-$k$ sparsification is an unbiased compression operator \cite{stich2018sparsified,wangni2018gradient}, it leads to larger compression errors and therefore has worse practical results compared to top-$k$ sparsification in the high compression regime \cite{eghlidi2020sparse, shi2019understanding}. Generally speaking, sparsification techniques reduce communication overload in a more aggressive manner compared to quantization. For example, the top-$k$ sparsification with error feedback can maintain the desired convergence rate and accuracy even with 99$\%$-99.9$\%$ gradient elements zeroed out \cite{aji2017sparse, lin2017deep}.

In contrast to sparsification in centralized learning, sparsification in federated learning has to consider not only the computation and storage efficiency but also the communication efficiency among distributed {{clients}} throughout the learning process. Therefore, most of the FL research on sparsification focus on improving existing sparsification techniques like top-$k$ sparsification to meet the needs for frequent communications better.

\textbf{Adaptive Sparsification.} In \citet{sahu2021rethinking}'s work, top-$k$ is proven to be the communication-optimal sparsifier with a given $k$ element budget per iteration from the optimization perspective, but a different hard-threshold sparsifier is further developed to consider optimality throughout the training instead of per iteration optimality. The proposed hard-threshold sparsifier adaptively determines the degree of sparsity $k$ by a constant hard threshold and has been proven to be the optimal sparsifier that theoretically minimizes the compression error under a given budget throughout the FL training process \cite{sahu2021rethinking}. Similarly, the adaptation of the sparsity degree $k$ 
to minimize the overall training time has been proposed \cite{9355797}. It is achieved by adjusting the degree of sparsity to come close to achieving the ideal balance between computation and communication in an online learning manner. 

Some other work on adaptive sparsification focuses not only on adjusting the sparsity level $d$ but also on {{co-adjustment}} with some other factors in federated learning, such as local update iterations and partial participation. For example, \citet{8852172} provides an information-theoretic way to analyze communication delay and error-accumulating sparsification techniques, both of which achieve information compression by delaying some updates and gathering gradient information before actual transmission. On the basis of these theoretical perspectives, the proposed Sparse Binary Compression (SBC) method adaptively trades off the temporal sparsity and gradient sparsity.  Similarly, \citet{nori2021fast} considers both local update and sparsity budgets to characterize learning error and adaptively adjusts these two components to yield Fast FL (FFL). {{As another line of work, \citet{abdelmoniem2021dc2} design an adaptive compression control mechanism that better trades-off between training speed and accuracy by coupling network delays and compression control. The superiority of this method is demonstrated with top-$k$ and rand-$k$.}}

To overcome the staleness of updates resulting from sparsified communication \citet{9148987} propose a General Gradient Sparsification (GGS) framework which performs gradient correction on the accumulated insignificant gradients for adaptive optimizers and updates the batch normalization layer with clients' local gradients. This method alleviates the impact of delayed gradient elements and thus further improves the performance of FL with sparsity. As another line of work, the linear FetchSGD sketch proposed by \citet{pmlr-v119-rothchild20a} allows error accumulation to be moved from local clients to the central server, eliminating the need for local client states. FetchSGD is an effective strategy to address the challenges of partial participation in FL systems. 

\textbf{Bidirectional Sparsification.} The aforementioned sparsification methods primarily consider compressing the upstream communication from clients to the central server. Due to the heterogeneity of local data, the sparsity patterns of updates from different clients can be very different. \citet{8889996} have claimed that if the amount of clients is larger than a threshold, the downstream update will be dense. To solve this problem, the sparse ternary compression (STC) framework is specifically designed to extend the top-$k$ sparsification for enabling downstream compression. \citet{8884924} introduce a global Top-k (gTop-k) sparsification method, which utilizes a tree structure to determine the global $k$ largest absolute values of all clients. This method overcomes the challenge of the irregular non-zero gradient indices from different clients during aggregation and, therefore, successfully compresses the downstream communication as well.

Another way to achieve bidirectional communication compression is to reduce the extra budget caused by specifying non-zero location indices. \citet{NEURIPS2021_b0ab42fc} develop a synergistic combination of various compressors for both gradient values and indices. The proposed Bloom-filter-based index compressor can reduce 50$\%$ of transmission compared against raw \textit{$\langle$key, value$\rangle$} sparse representation. Besides, it is also effective to put limitations on transmitting new non-zero positions \cite{ozfatura2021time}. In this work, the proposed time-correlation sparsification (TCS) scheme utilizes the correlation between consecutive sparse representations in FL training to reduce the transmission of newly-computed non-zero indices. This method has been shown to achieve a higher compression level in downstream communication compared to upstream communication.

\textbf{Layer-wise Sparsification with Pipeline.} With the aim of further improving the communication efficiency of FL in terms of accelerating the entire training process, some works on FL with sparsity have considered making use of the models' layered structure to parallelize the communications with computations, which is referred to the pipeline \cite{li2018pipe, harlap2018pipedream, shi2019mg}. To combine gradient sparsification with the idea of the pipeline, \citet{shi2019layer} develop a layer-wise adaptive gradient sparsification (LAGS) scheme, where the subset of remaining values for each layer is selected independently according to a given ratio, and the transmission of any sparsified gradient of any layer $l+1$' can be parallelized with the gradient calculation and sparsification of layer $l$, instead of waiting for the completion of the entire back-propagation before transmitting a single sparsified gradient.

However, an existing drawback of LAGS is that the sparsified gradient of each layer will invoke one independent communication, and thus the layer-wise communications require high communication start-up costs \cite{shi2019layer, you2017scaling}.  One intensively-studied way to alleviate the startup overload is merging gradients from multiple layers for one communication, but it will also cause more computation and waiting time \cite{shi2019layer, you2017scaling, sergeev2018horovod, jia2018highly}. To further trade-off gradient computation and layer-wise gradient sparsification, an optimal merging scheme named Optimal Merged Gradient Sparsification (OMGS) has been developed \cite{9155269}. It formulates the trade-off as an optimization problem and minimizes the iteration time by maximizing the overlap between sparsified gradient computation and communication during training.  


\subsection{Distillation}
Knowledge distillation (KD) \cite{hinton2015distilling} is further employed to FL to alleviate the communication bottleneck. Precisely, in order to reduce the model size, KD aims to transfer the information of a large model (teacher/mentor) to a small model (student/mentee) without adversely impacting the model's precision and convergence. Therefore, the teacher network is typically a network with a large number of parameters and a complicated structure, with excellent performance and generalization ability, whereas the student network has a modest number of parameters and a simple structure. Since the shape sizes of different student models' outputs are identical and irrelevant to the structure of student models, KD could be utilized to tackle the model heterogeneity issue as aforementioned by uploading the output of the local model without softmax (i.e., the logit vector) other than the model itself, and such method is called federated distillation (FD). In this section, we give a taxonomy of FD based on whether the client will upload a small subset of their private data to build a public dataset on the {{server side}} and divide FD into the data-additional FD (need to construct an addition dataset {{jointly}}) methods and the data-free FD methods.

\textbf{Data-additional FD.}
At first, \citet{li2019fedmd} propose FedMD that each client first trains the local models on a labeled public dataset and uploads their class scores (i.e., logit vector) rather than model {{parameters}} to the server, and subsequently, the server integrates them to obtain the knowledge from all clients. However, the training process of FedMD is simultaneously on both labeled public data and private data, which {{necessitates a considerable amount of local computation}} and is not suitable for resource-limited devices. Furthermore, the creation of public datasets necessitates careful deliberation and thus lacks generalization.
Consequently, \citet{lin2020ensemble} propose FedDF to train the global model through an unlabeled dataset in an ensemble distillation manner and prove that FedDF is robust to the selection and combination of the FD dataset. Especially, the selection of an auxiliary dataset is set on the {{server side}} and effectively mitigates the computational pressure of local clients. Since the public dataset is unlabeled, the privacy leakage of FedDF is also much less than the FedMD's. Nonetheless, although the selection of the public dataset is moved to the {{server side}}, there are still no criteria for determining the size of the dataset.
In \citet{sattler2021cfd}, they establish the Compressed Federated Distillation (CFD) developing {{``entropy"}}, ``certainty", and ``margin" standards to determine the size of the public distillation dataset. Afterward, they also utilize a uniform quantization algorithm and delta coding to reduce the communication cost further.

The previous work mainly concentrates on reducing the communication cost of FD by either changing the size of the public dataset or the size of logit vectors, but none of them considers modifying the \textbf{aggregation strategy}.
By contrast, \cite{itahara2021distillation} propose DS-FL, which attempts to aggregate the local models via an entropy reduction strategy. Since the existence of data heterogeneity in FL, the entropy of global logits is quite high, which indicates the difficulty of training a {{model with}} high performance in {{non-iid}} scenarios. DS-FL reduces the entropy of global logits sharpening logits by adding the temperature factor $T$ to the softmax to accelerate and stabilize the DS-FL. In addition, another aggregation algorithm named FedAUX is created by \citet{sattler2021fedaux}, which calculates the certainty of each client's logits by contrastive logistic scoring and uses this certainty {{to determine}} the aggregation {{weights}} to yield a strong empirical performance on data with high distinction. Moreover, \citet{li2021personalized} construct an adaptive aggregation strategy named pFedSD to dynamically modify the weight of each participant and improve the quality and performance of the global model. Specifically, it exploits the Jensen-Shannon divergence between the transmitted model outputs in two adjacent rounds as the weight of clients to measure the similarity of local models. Similarly, \citet{sturluson2021fedrad} also develop an adaptive aggregator named FedRAD based on median scores to reinforce the global model.

Combining both public dataset selection and non-trivial aggregation methods, \citet{liu2022communication} present FAS to actively {{sample}} client data to ameliorate the convergence of the training process under client-drift situations and provide its theoretical analysis. Specifically, they solve the {{non-iid}} problem by weighted logistic scores based on the discrepancy of clients' data and then add the differential privacy mechanism to the aggregation algorithm to preserve the safety of the local private data coupled with restricting the privacy loss.

{{\textbf{Data-free FD.}}}
Although in the data-additional FD category, transferring logits can greatly reduce the communication overhead, it requires each client to sacrifice a part of its own private data to construct a public dataset, which might be unacceptable in some extremely private cases, such as a user's private information data on their personal devices.
Therefore, data-free FD algorithms are proposed without sharing any private data to further preserve the security of all clients. The first data-free FD algorithm is developed by \citet{jeong2018communication} to handle the {{non-iid}} problem, in which clients periodically send their {{average}} local logits to the server and afterward receive the globally aggregated logit as the teacher's output to help train the local student model. However, the issue of model heterogeneity is not addressed in this research. To this end, \citet{jiang2020federated} present FedDistill and solve this problem by forcing the client to share a global model of the same structure while training another customized local model. Concretely, each client holds two models: a personalized large model that serves as the teacher model and is trained on local data, and a small global model received from the server. In each iteration, student models with the same structure are transmitted to the server and aggregated, which could still obtain all clients' information with arbitrary personalized large model structures. Following FedDistill's lead, \citet{wu2022communication} introduce FedKD including three different loss functions to encourage the training of the global model and matrix factorization to further reduce the communication overhead. Specifically, they utilize the Kullback–Leibler (KL) divergence loss to adaptively distill the knowledge of both teacher and student models.

Nevertheless, directly aggregating the global model might still be heavily affected by the heterogeneity issue, and its performance might be unsatisfactory. Thus, \citet{zhu2021data} propose FedGEN to train a \textbf{generator} on the {{server side}} based on the transmitted classifiers received from clients. Specifically, in each FL iteration, clients convey their local label count information to the server and regularize the local training after obtaining the global lightweight generator.

{{After the release of FedGEN for training a generator to facilitate the FD procedure, some other recent work focuses on improving the \textbf{performance} of the generator model. For example, \citet{yao2021local} propose FedGKD to utilize the averaged parameters of historical global models for the ensemble, which mitigate the client-drift problem caused by the non-iid local dataset. In addition, \citet{zhang2022fine} propose FedFTG that first generates pseudo-data to train the generator and then utilizes the hard samples to train the global model simultaneously. What's more, facing the data distribution shift issue, they also devise customized label sampling and label-wise ensemble algorithms to boost the convergence of the FD process.}}

\textbf{Applications.}
{{Thanks to its special teacher-student fashion, other than employed to communication-efficient scenarios, KD is also frequently used in addressing the heterogeneity issues of clients, which is also called \textit{personalized FL}. 
\citet{cho2022heterogeneous} propose a weighted consensus KD framework dubbed Fed-ET, which utilizes consensus distillation with diversity regularization to better extract the model knowledge based on heterogeneous scenarios.
Additionally, \citet{zhu2022resilient} devise a personalized framework FedResCuE, which trains a number of sub-models by pruning the full global model corresponding to a pruning sequence. In this way, each client just selects the pruning weight closing to its own local model and only receives a tiny model rather than the cumbersome global model.
Besides, \citet{zhang2022fedzkt} introduce a knowledge agnostic KD framework (even without any prior knowledge) called FedZKT transmitting a ``personalized'' generator based on the local model structure of each client and reducing the communication cost further. Furthermore, \citet{ozkara2021quped} develop a compressed personalized KD algorithm called QuPeD, which applies a soft quantization method by solving an optimization problem and utilizes KD to tackle the resource heterogeneity issue.}}

On account of the impressive feasibility and compression ability, KD has been wildly employed in massive FL scenarios. For instance, due to the high communication and computation overhead in FL protection methods such as local differential privacy (LDP) \cite{kairouz2014extremal} and secure multi-party computation (SMPC) \cite{du2001secure}, it is complicated to incorporate these secure compute methods into real-world FL applications. Therefore, a variety of current research \cite{sun2020federated, oh2020mix2fld, shi2021towards, wu2022federated, chang2019cronus, cha2019federated, cha2020proxy} considers combining KD with secure computation to boost the whole training process and privacy-preserving medical prediction \cite{sui2020feded, sun2022fed2kd}. Moreover, KD is also used in IoT and edge learning \cite{ahn2019wireless, ahn2020cooperative, wang2020industrial} to reduce communication costs.

\subsection{Minors}

\textbf{Reduce Communication Rounds / Fast Convergence.}
In this article that proposes federated learning \cite{mcmahan2017communication}, they propose FedAvg to perform global aggregation after multiple iterations of local updates rather than directly collecting the model gradient in each iteration, which hugely reduces the communication frequency. Moreover, \citet{li2019fair} introduce q-FedAvg to dynamically select a subset of all clients and achieve faster convergence in terms of communication rounds. Subsequently, \citet{hyeon2021fedpara} boost the convergence rate via low-rank Hadamard product parameterization. Furthermore, \cite{sun2020lazily} propose a criteria framework named LAQ to only update the compressed model gradient when the variation of the local model is large enough, and thus reduce the number of communication rounds.

\textbf{Low-rank Decomposition.}
\citet{konevcny2016federated} consider {{uploading}} structured models in FL and propose modifying the updated matrix to a sparse low-rank matrix to save communication cost, allowing the low-rank approximation in FL communication. Furthermore, \citet{wu2022communication} discover that the updated gradients have low-rank properties and thus utilize the singular value decomposition (SVD) \cite{wall2003singular} to decompose the model gradient matrix into smaller matrices, which significantly mitigates communication overhead. {{Additionally, \citet{azam2021recycling} propose to recycle the gradients between communication rounds by utilizing the low-rank property, which reduces the transmission of model parameters to single scalars.}}

\textbf{Topology Design}
{{Unlike the client-server architecture we mainly concentrate on, the topological network design for cross-silo FL \cite{kairouz2021advances} is also an area of interest.
\citet{marfoq2020throughput} claim that the high-speed access links are more efficient in exchanging information in cross-silo scenarios and introduce a novel topology design by optimizing the minimal cycle time problem and obtaining the largest throughput.
Furthermore, \citet{guo2022hybrid} propose HL-SGD utilizing the device-to-device(D2D) communication capabilities to divide clients into a set of separate clusters with high D2D communication bandwidth and speeding global model convergence, without losing the model performance.}}

\subsection{Comparison and discussion}
{{In these aforementioned subsections, we mainly introduce three kinds of communication-efficient strategies. However, each of them has its own merits and demerits. For quantization methods, since most of the time quantizers are defined explicitly, the convergence analysis can be performed directly on them. Moreover, sparsification is more empirical and sometimes can mitigate more communication costs than quantization methods. Nevertheless, we cannot provide some theoretical analysis of this method. For knowledge distillation, it has an excellent performance in heterogeneous FL scenarios. But similar to sparsification, no theoretical analysis we can derive.}}

\setlength{\tabcolsep}{0mm}
\setlength\LTleft{-1in}
\setlength\LTright{-1in plus 1 fill}
{
\small
\tiny

\begin{longtable}[c]{ccccccccccc}
\caption{Classification and comparison of surveyed FL work on communication efficiency. \textit{Amount/Round} means the strategy reduces the amount of communication or the rounds. \textit{Partial node} indicates whether the mentioned method supports the situation that some clients participate while some clients drop the line. \textit{Down/Up} implies the effectiveness of the cited method is on the downstream or the upstream.}
\label{Table1}\\
\hline
\toprule[0.8pt]
 &  &  & \multicolumn{2}{c}{\textbf{Method}} & \multicolumn{3}{c}{\textbf{Communication}} & \multicolumn{3}{c}{\textbf{Evaluation}} \\ \cline{4-11} 
\multirow{-2}{*}{\textbf{}} & \multirow{-2}{*}{\textbf{\begin{tabular}[c]{@{}c@{}}Compression\\ level\end{tabular}}} & \multirow{-2}{*}{\textbf{Ref.}} & \textbf{\begin{tabular}[c]{@{}c@{}}Theoretical \\ guarantee\end{tabular}} & \textbf{non-iid} & \textbf{\begin{tabular}[c]{@{}c@{}}Amount/\\ Round\end{tabular}} & \textbf{\begin{tabular}[c]{@{}c@{}}Partial\\ node\end{tabular}} & \textbf{\begin{tabular}[c]{@{}c@{}}Down/\\ Up\end{tabular}} & \textbf{Datasets} & \textbf{\begin{tabular}[c]{@{}c@{}}\# of \\ devices\end{tabular}} & \textbf{\begin{tabular}[c]{@{}c@{}}FL\\ baselines\end{tabular}} \\ \hline
\endfirsthead
\endhead
\toprule[0.8pt]
{\color[HTML]{F8A102} } &  & \cite{reisizadeh2020fedpaq} & \CheckmarkBold & \XSolidBrush & Amount & \CheckmarkBold & Both & \begin{tabular}[c]{@{}c@{}}MNIST,\\ CIFAR-10\end{tabular} & 50 & QSGD \\ \cline{3-11} 
{\color[HTML]{F8A102} } &  & \cite{amiri2020federated2} & \CheckmarkBold & \CheckmarkBold & Amount & \XSolidBrush & U & MNIST & 25 & \begin{tabular}[c]{@{}c@{}}SignSGD,\\ QSGD,\\ D-DSGD\end{tabular} \\ \cline{3-11} 
{\color[HTML]{F8A102} } &  & \cite{das2020faster} & \CheckmarkBold & \CheckmarkBold & Amount & \XSolidBrush & U & \begin{tabular}[c]{@{}c@{}}CIFAR10,\\ FMNIST\end{tabular} & 50 & FedAvg,FedPAQ \\ \cline{3-11} 
{\color[HTML]{F8A102} } & \multirow{-4}{*}{\textbf{mid}} & \cite{dai2019hyper} & \CheckmarkBold & \XSolidBrush & Both & \CheckmarkBold & U & \begin{tabular}[c]{@{}c@{}}ILSVRC-12, \\ CIFAR-10,\\ CIFAR-100\end{tabular} & 1000, 10\% & \begin{tabular}[c]{@{}c@{}}QSGD,\\ TernGrad,\\ SignSGD,\\ SGD\end{tabular} \\ \cline{2-11} 
{\color[HTML]{F8A102} } &  & \cite{jhunjhunwala2021adaptive} & \CheckmarkBold & \CheckmarkBold & Both & \XSolidBrush & U & \begin{tabular}[c]{@{}c@{}}CIFAR-10,\\ FMNIST\end{tabular} & 4, 8 & NULL \\ \cline{3-11} 
\multirow{-9}{*}{{\color[HTML]{F8A102} \textbf{\rotatebox{90}{Stochastic Quantzation}}}} & \multirow{-2}{*}{\textbf{high}} & \cite{haddadpour2021federated} & \CheckmarkBold & \CheckmarkBold & Amount & \CheckmarkBold & U & \begin{tabular}[c]{@{}c@{}}MNIST,\\ CIFAR-10,\\ FMNIST,\\ EMNIST\end{tabular} & 100 & \begin{tabular}[c]{@{}c@{}}FedAvg, \\ FedPAQ,  \\ SCAFFOLD\end{tabular} \\ \hline \toprule[0.8pt]
{\color[HTML]{F8A102} } & \textbf{low} & \cite{khalfoun2021eden} & \XSolidBrush & \XSolidBrush & Both & \XSolidBrush & U & \begin{tabular}[c]{@{}c@{}}Geolife,\\ MDC,\\ Privamov\end{tabular} & 42, 144,48 & \begin{tabular}[c]{@{}c@{}}Geoi,\\ TRL,\\ PROM\end{tabular} \\ \cline{2-11}
{\color[HTML]{F8A102} } & \textbf{mid} & \cite{basat2022quick} & \CheckmarkBold & \XSolidBrush & Both & \XSolidBrush & Both & \begin{tabular}[c]{@{}c@{}}CIFAR-10, \\ CIFAR-100, \\ Shakespeare\end{tabular} & 50, 50, 10 & \begin{tabular}[c]{@{}c@{}}Hadamard, \\ QSGD, \\ EDEN\end{tabular} \\ \cline{2-11} 
\multirow{-9}{*}{{\color[HTML]{F8A102} \textbf{\rotatebox{90}{Rotation-based Quantization}}}} & \textbf{high} & \cite{vargaftik2021drive} & \CheckmarkBold & \XSolidBrush & Amount & \XSolidBrush & U & \begin{tabular}[c]{@{}c@{}}MNIST, \\ EMNIST, \\ CIFAR-10,\\ Shakespeare,\\ Stack Overflow\end{tabular} & 10 & \begin{tabular}[c]{@{}c@{}}FedAvg,\\ TernGrad,\\ 1-bit SQ\end{tabular} \\ \hline \toprule[0.8pt]
{\color[HTML]{F8A102} } & \textbf{low} & \cite{shlezinger2020federated} & \XSolidBrush & \XSolidBrush & Amount & \XSolidBrush & U & \begin{tabular}[c]{@{}c@{}}Gaussian iid \\ matrix\end{tabular} & \begin{tabular}[c]{@{}c@{}}from 3\\ to 15\end{tabular} & \begin{tabular}[c]{@{}c@{}}Uniform\\ quantizer\end{tabular} \\ \cline{2-11} 
{\color[HTML]{F8A102} } &  & \cite{shlezinger2020uveqfed} & \CheckmarkBold & \XSolidBrush & Amount & \XSolidBrush & U & \begin{tabular}[c]{@{}c@{}}MNIST,\\ CIFAR-10\end{tabular} & 100/15, 10 & \begin{tabular}[c]{@{}c@{}}FedAvg,\\ QSGD,\\ Uniform \\ quantizer\end{tabular} \\ \cline{3-11} 
{\color[HTML]{F8A102} } & \multirow{-4}{*}{\textbf{mid}} & \cite{chen2021communication} & \CheckmarkBold & \XSolidBrush & Both & \XSolidBrush & U & \begin{tabular}[c]{@{}c@{}}MNIST,\\ Finger \\ Movement\end{tabular} & 15 & \begin{tabular}[c]{@{}c@{}}QSGD,\\ Uniform \\ quantizer\end{tabular} \\ \cline{2-11} 
{\color[HTML]{F8A102} } &  & \cite{jin2020stochastic} & \CheckmarkBold & \CheckmarkBold & Amount & \XSolidBrush & Both & \begin{tabular}[c]{@{}c@{}}MNIST, \\ CIFAR-10\end{tabular} & 31 & SignSGD \\ \cline{3-11} 
\multirow{-12}{*}{{\color[HTML]{F8A102} \textbf{\rotatebox{90}{Lattice-based Quantization}}}} & \multirow{-2}{*}{\textbf{high}} & \cite{zhu2020one} & \CheckmarkBold & \XSolidBrush & Amount & \XSolidBrush & Both & \begin{tabular}[c]{@{}c@{}}MNIST,\\ CIFAR-10\end{tabular} & 100 & BAA \\ \hline \toprule[0.8pt]
{\color[HTML]{F8A102} } & \textbf{low} & \cite{li2021communication} & \XSolidBrush & \CheckmarkBold & Both & \CheckmarkBold & Both & \begin{tabular}[c]{@{}c@{}}MNIST,\\ FMNIST\end{tabular} & 10 & \begin{tabular}[c]{@{}c@{}}FedAvg,\\ SignSGD\end{tabular} \\ \cline{2-11} 
{\color[HTML]{F8A102} } & \textbf{mid} & \cite{abdi2020quantized} & \CheckmarkBold & \XSolidBrush & Amount & \CheckmarkBold & U & \begin{tabular}[c]{@{}c@{}}CIFAR-10,\\ ImageNet\end{tabular} & \begin{tabular}[c]{@{}c@{}}from 2\\ to 16\end{tabular} & \begin{tabular}[c]{@{}c@{}}Random k, \\ Top k, \\ Threshold \\ TernGrad, \\ Adaptive \\ Threshold\end{tabular} \\ \cline{2-11} 
{\color[HTML]{F8A102} } &  & \cite{fan2021communication} & \CheckmarkBold & \XSolidBrush & Amount & \XSolidBrush & Both & MNIST & 10 & NULL \\ \cline{3-11} 
\multirow{-8}{*}{{\color[HTML]{F8A102} \textbf{\rotatebox{90}{\begin{tabular}[c]{@{}c@{}}Quantized\\ Compressed Sensing\end{tabular}}}}} & \multirow{-2}{*}{\textbf{high}} & \cite{oh2021quantized} & \XSolidBrush & \CheckmarkBold & Amount & \XSolidBrush & Both & MNIST & 30 & \begin{tabular}[c]{@{}c@{}}QCS-QIHT,\\ QCS-Dither,\\ SignSGD\end{tabular} \\ \hline \toprule[0.8pt]
{\color[HTML]{57DD56} } & \textbf{low} & \cite{nori2021fast} & \CheckmarkBold & \CheckmarkBold & Both & \CheckmarkBold & Both & \begin{tabular}[c]{@{}c@{}}MNIST,\\      CIFAR-10\end{tabular} & 32, 16 & \begin{tabular}[c]{@{}c@{}}ADACOMM,\\ ATOMO\end{tabular} \\ \cline{2-11} 
{\color[HTML]{57DD56} } & \textbf{mid} & \cite{9355797} & \CheckmarkBold & \CheckmarkBold & Both & \XSolidBrush & Both & \begin{tabular}[c]{@{}c@{}}FEMNIST,\\ CIFAR-10\end{tabular} & 156, 100 & \begin{tabular}[c]{@{}c@{}}Periodic-k,\\ Top-k,\\ FedAvg\end{tabular} \\ \cline{2-11}
\multirow{-5}{*}{{\color[HTML]{57DD56} \textbf{\rotatebox{90}{\begin{tabular}[c]{@{}c@{}}Adaptive \\ Sparsification\end{tabular}}}}} & \multirow{1}{*}{\textbf{high}}& \cite{8852172} & \CheckmarkBold & \XSolidBrush & Amount & \XSolidBrush & U & \begin{tabular}[c]{@{}c@{}}MNIST,\\ CIFAR-10,\\ CIFAR-100,\\ ImageNet,\\ PTB,\\ Wikitext-2\end{tabular} & \begin{tabular}[c]{@{}c@{}}from 4\\ to 400\end{tabular} & FedAVG \\ \cline{3-11} 
 {\color[HTML]{57DD56} } & & \cite{9148987} & \CheckmarkBold & \XSolidBrush & Amount & \XSolidBrush & U & \begin{tabular}[c]{@{}c@{}}MNIST,\\ CIFAR-10,\\ ImageNet\end{tabular} & 2, 4, 8 & TernGrad \\ \hline \toprule[0.8pt]
{\color[HTML]{57DD56} } &  & \cite{8889996} & \CheckmarkBold & \CheckmarkBold & Amount & \CheckmarkBold & Both & \begin{tabular}[c]{@{}c@{}}CIFAR-10,\\ MNIST,\\ FEMNIST,\\ KWS\end{tabular} & 100, 10\% & \begin{tabular}[c]{@{}c@{}}FedAvg,\\ SignSGD\end{tabular} \\ \cline{3-11} 
{\color[HTML]{57DD56} } &  & \cite{8884924} & \XSolidBrush & \XSolidBrush & Amount & \XSolidBrush & U & \begin{tabular}[c]{@{}c@{}}CIFAR-10, \\ ImageNet,\\ Penn \\ Treebank\\ corpus\end{tabular} & \begin{tabular}[c]{@{}c@{}}from 4\\ to 128\end{tabular} & \begin{tabular}[c]{@{}c@{}}S-SGD,\\ Top-k\end{tabular} \\ \cline{3-11} 
{\color[HTML]{57DD56} } & \multirow{-9}{*}{\textbf{mid}} & \cite{NEURIPS2021_b0ab42fc} & \CheckmarkBold & \XSolidBrush & Amount & \XSolidBrush & Both & \begin{tabular}[c]{@{}c@{}}CIFAR-10,\\ ImageNet, \\ Stack Overflow\end{tabular} & 56 & \begin{tabular}[c]{@{}c@{}}FedAvg,\\ Top-k\end{tabular} \\ \cline{2-11} 
\multirow{-13}{*}{{\color[HTML]{57DD56} \textbf{\rotatebox{90}{\begin{tabular}[c]{@{}c@{}}Bidirectional \\ Sparsification\end{tabular}}}}} & \textbf{high} & \cite{ozfatura2021time} & \XSolidBrush & \XSolidBrush & Amount & \XSolidBrush & Both & CIFAR-10 & 10 & Top-k \\ \hline \toprule[0.8pt]
{\color[HTML]{57DD56} } & \textbf{low} & \cite{9155269} & \CheckmarkBold & \XSolidBrush & Both & \XSolidBrush & U & \begin{tabular}[c]{@{}c@{}}CIFAR-10,\\ ImageNet,\\ PTB\end{tabular} & 16 & \begin{tabular}[c]{@{}c@{}}MGS-SGD,\\ Topk-SGD,\\ P-SGD,\\ LAGS-SGD\end{tabular} \\ \cline{2-11} 
\multirow{-5}{*}{{\color[HTML]{57DD56} \textbf{\rotatebox{90}{\begin{tabular}[c]{@{}c@{}}Layer-wise\\sparsification \end{tabular} }}}} & \textbf{mid} & \cite{shi2019layer} & \CheckmarkBold & \XSolidBrush & Amount & \XSolidBrush & U & \begin{tabular}[c]{@{}c@{}}CIFAR-10,\\ Inception-v4,\\ ImageNet, \\ PTB\end{tabular} & 16 & \begin{tabular}[c]{@{}c@{}}SLGS-SGD,\\ Dense-SGD.\end{tabular} \\ \hline \toprule[0.8pt]
{\color[HTML]{34CDF9} } &  & \cite{li2019fedmd} & \XSolidBrush & \CheckmarkBold & Amount & \XSolidBrush & D & \begin{tabular}[c]{@{}c@{}}MNIST,\\ FMNIST\\ CIFAR-10,\\ CIFAR-100\end{tabular} & NULL & NULL \\ \cline{3-11} 
\multirow{8}{*}{{\color[HTML]{34CDF9} \textbf{\rotatebox{90}{Data-additional FD}}}} & \multirow{-5}{*}{\textbf{low}} & \cite{sturluson2021fedrad} & \XSolidBrush & \CheckmarkBold & Amount & \XSolidBrush & U & MNIST & 30 & \begin{tabular}[c]{@{}c@{}}FedAvg, \\ AFA,\\ FedMGDA, \\ FedDF\end{tabular} \\ \cline{2-11} 
{\color[HTML]{34CDF9} } &  & \cite{lin2020ensemble} & \XSolidBrush & \CheckmarkBold & Amount & \CheckmarkBold & U & \begin{tabular}[c]{@{}c@{}}CIFAR-10,\\ CIFAR-100, \\ ImageNet, \\ AG News, \\ SST2\end{tabular} & 50, 40\% & \begin{tabular}[c]{@{}c@{}}FedMD,\\ FedAvg, \\ FedProx\end{tabular} \\ \cline{3-11} 
{\color[HTML]{34CDF9} } &  & \cite{liu2022communication} & \XSolidBrush & \CheckmarkBold & Both & \XSolidBrush & Both & \begin{tabular}[c]{@{}c@{}}STL-10, \\ CIFAR-10\end{tabular} & 20 & \begin{tabular}[c]{@{}c@{}}FedAvg,\\ FedDF\end{tabular} \\ \cline{3-11} 
{\color[HTML]{34CDF9} } &  & \cite{itahara2021distillation} & \XSolidBrush & \CheckmarkBold & Amount & \XSolidBrush & Both & \begin{tabular}[c]{@{}c@{}}MNIST, \\ FMNIST\end{tabular} & 0 & \begin{tabular}[c]{@{}c@{}}FedAvg,\\ FD\end{tabular} \\ \cline{3-11} 
{\color[HTML]{34CDF9} } & \multirow{-7}{*}{\textbf{mid}} & \cite{li2021personalized} & \XSolidBrush & \CheckmarkBold & Both & \CheckmarkBold & U & \begin{tabular}[c]{@{}c@{}}MNIST, \\ FMNIST, \\ CIFAR-10\end{tabular} & 10, 80\% & \begin{tabular}[c]{@{}c@{}}FedMD,\\ DS-FL,\\ MHAT\end{tabular} \\ \cline{2-11} 
{\color[HTML]{34CDF9} } &  & \cite{sattler2021cfd} & \XSolidBrush & \CheckmarkBold & Both & \XSolidBrush & Both & \begin{tabular}[c]{@{}c@{}}MNIST, \\ EMNIST, \\ CIFAR-10, \\ STL-10\end{tabular} & 20 & \begin{tabular}[c]{@{}c@{}}FedAvg,\\ FD\end{tabular} \\ \cline{3-11} 
{\color[HTML]{34CDF9} } & \multirow{-5}{*}{\textbf{high}} & \cite{sattler2021fedaux} & \XSolidBrush & \CheckmarkBold & Both & \CheckmarkBold & U & \begin{tabular}[c]{@{}c@{}}STL-10, \\ CIFAR-100, \\ SVHN,\\ ImageNet\end{tabular} & 100 & \begin{tabular}[c]{@{}c@{}}FedDF,\\ FedAvg\end{tabular} \\ \hline \toprule[0.8pt]
{\color[HTML]{34CDF9} } &  & \cite{jeong2018communication} & \XSolidBrush & \CheckmarkBold & Amount & \XSolidBrush & Both & MNIST & NULL & FAug \\ \cline{3-11} 
{\color[HTML]{34CDF9} } & \multirow{-2}{*}{\textbf{mid}} & \cite{jiang2020federated} & \XSolidBrush & \CheckmarkBold & Amount & \XSolidBrush & U & \begin{tabular}[c]{@{}c@{}}MNIST, \\ CIFAR-10\end{tabular} & NULL & FedAvg \\ \cline{2-11} 
\multirow{0}{*}{{\color[HTML]{34CDF9} \textbf{\rotatebox{90}{Data-free FD}}}} & \multirow{13}{*}{\textbf{high}}  & \cite{zhu2021data} & \XSolidBrush & \CheckmarkBold & Both & \CheckmarkBold & Both & \begin{tabular}[c]{@{}c@{}}MNIST, \\ EMNIST, \\ CELEBA\end{tabular} & 20, 50\% & \begin{tabular}[c]{@{}c@{}}FedAvg, \\ FedProx,\\ FedEnsemble,\\ FedDfusion,\\ FedDistill\end{tabular} \\ \cline{3-11} 
{\color[HTML]{34CDF9} } &  & \cite{yao2021local} & \CheckmarkBold & \CheckmarkBold & Both & \CheckmarkBold & Both & \begin{tabular}[c]{@{}c@{}}CIFAR-10,\\      CIFAR-100, \\ ImageNet, \\ AG News, \\ SST-5\end{tabular} & NULL & \begin{tabular}[c]{@{}c@{}}FedAvg,\\ FedProx,\\ MOON,\\ FedDistill,\\ FedGen\end{tabular} \\ \cline{3-11} 
{\color[HTML]{34CDF9} } & & \cite{zhang2022fine} & \XSolidBrush & \CheckmarkBold & Both & \CheckmarkBold & Both & \begin{tabular}[c]{@{}c@{}}CIFAR-10, \\ CIFAR-100\end{tabular} & 100, 10\% & \begin{tabular}[c]{@{}c@{}}FedAvg,\\ FedProx,\\ SCAFFOLD,\\ FedDyn,\\ MOON,\\ FedGen,\\ FedDF\end{tabular} \\ \hline \toprule[0.8pt]
{\color[HTML]{6665CD} } &  & \cite{mcmahan2017communication} & \XSolidBrush & \CheckmarkBold & Rounds & \CheckmarkBold & U & \begin{tabular}[c]{@{}c@{}}MNIST,\\ Shakespeare\end{tabular} & 100,10\% & NULL \\
{\color[HTML]{6665CD} } & \multirow{-2}{*}{\textbf{low}} & \cite{li2019fair} & \XSolidBrush & \CheckmarkBold & Rounds & \CheckmarkBold & U & \begin{tabular}[c]{@{}c@{}}Adult,\\ FMNIST\end{tabular} & \begin{tabular}[c]{@{}c@{}}from 23 \\ to 1101\end{tabular} & AFL \\ \cline{2-11} 
{\color[HTML]{6665CD} } & \textbf{mid} & \cite{hyeon2021fedpara} & \XSolidBrush & \CheckmarkBold & Rounds & \CheckmarkBold & U & \begin{tabular}[c]{@{}c@{}}CIFAR-10,\\ CIFAR-100,\\ CINIC-10,\\ Shakespeare\end{tabular} & 100,16\% & \begin{tabular}[c]{@{}c@{}}FedAvg,\\ FedProx,\\ SCAFFOLD\end{tabular} \\ \cline{2-11} 
\multirow{-9}{*}{{\color[HTML]{6665CD} \textbf{\rotatebox{90}{Reduce rounds}}}} & \textbf{high} & \cite{sun2020lazily} & \CheckmarkBold & \XSolidBrush & Both & \XSolidBrush & Both & \begin{tabular}[c]{@{}c@{}}MNIST,\\ ijcnn1,\\ covtype\end{tabular} & 10 & GD,QGD,LAG \\ \hline \toprule[0.8pt]
{\color[HTML]{6665CD} } & \textbf{mid} & \cite{konevcny2016federated} & \XSolidBrush & \XSolidBrush & Amount & \CheckmarkBold & U & \begin{tabular}[c]{@{}c@{}}CIFAR-10,\\ Reddit\\ dataset\end{tabular} & 100 & NULL \\ \cline{2-11} 
\multirow{-5}{*}{{\color[HTML]{6665CD} \textbf{\rotatebox{90}{\begin{tabular}[c]{@{}c@{}}Low-rank\\ Approximation\end{tabular}}}}} & \textbf{high} & \cite{wu2022communication} & \XSolidBrush & \CheckmarkBold & Amount & \XSolidBrush & Both & \begin{tabular}[c]{@{}c@{}}MIND, \\ ADR, \\ CADEC, \\ ADE, \\ SMM4H\end{tabular} & 1, 2, 3, 4 & \begin{tabular}[c]{@{}c@{}}FetchSGD, \\ FedDropout, \\ SCAFFOLD,\\ FedPAQ\end{tabular} \\ \hline \toprule[0.8pt]

\end{longtable}
}

\section{Communication Environment}

In the previous section, most of the papers assume that the communication channel is perfect and do not consider the resources of clients, such as the devices' energy and bandwidth. Furthermore, although the previous work proposes to compress the updated model or reduce the communication rounds to reduce the overall communication time, none of them study the problem under a time-limited scenario. In this section, we discuss the impact of different communication environments on model performance and convergence rate in FL and summarize the comparison of mentioned methods in \Cref{Table2}.

\begin{table}[]
\centering
\caption{Classification and comparison of surveyed FL work on the communication environment. \textit{Client selection} indicates whether this method performs the client selection process. \textit{Limited bandwidth} implies whether this strategy takes the limited bandwidth {{constraints}} into account.}
\label{Table2}
\resizebox{\columnwidth}{!}{%
\begin{tabular}{ccccccccccc}
\hline
\toprule[0.8pt]
\multirow{2}{*}{\textbf{}} & \multirow{2}{*}{\textbf{\begin{tabular}[c]{@{}c@{}}Down/\\ Up\end{tabular}}} & \multirow{2}{*}{\textbf{Ref.}} & \multicolumn{2}{c}{\textbf{Methods}} & \multicolumn{3}{c}{\textbf{Communication}} & \multicolumn{3}{c}{\textbf{Evaluation}} \\ \cline{4-11} 
 &  &  & \textbf{\begin{tabular}[c]{@{}c@{}}Theoretical\\ guarantee\end{tabular}} & \textbf{non-iid} & \textbf{\begin{tabular}[c]{@{}c@{}}Client\\ selection\end{tabular}} & \textbf{\begin{tabular}[c]{@{}c@{}}Limited\\ bandwidth\end{tabular}} & \textbf{\begin{tabular}[c]{@{}c@{}}Partial\\ node\end{tabular}} & \textbf{Datasets} & \textbf{\begin{tabular}[c]{@{}c@{}}\# of \\ devices\end{tabular}} & \textbf{\begin{tabular}[c]{@{}c@{}}FL\\ baselines\end{tabular}} \\ \hline \toprule[0.8pt]
 &  & \cite{wu2020safa} & \XSolidBrush & \CheckmarkBold & \CheckmarkBold & \CheckmarkBold & \CheckmarkBold & \begin{tabular}[c]{@{}c@{}}Boston \\ Housing, \\ MNIST, \\ KDD Cup’99\end{tabular} & 5, 100, 500 & \begin{tabular}[c]{@{}c@{}}FedAvg, \\ FedCS\end{tabular} \\ \cline{3-11} 
\multirow{5}{*}{\rotatebox{90}{\textbf{\begin{tabular}[c]{@{}c@{}} Unreliable\\ Network\end{tabular}}}} & \multirow{-3}{*}{\textbf{Down}} & \cite{salehi2021federated} & \CheckmarkBold & \CheckmarkBold & \CheckmarkBold & \XSolidBrush & \CheckmarkBold & MNIST & 100 & FedAvg \\ \cline{3-11} 
 &  & \cite{wu2020accelerating} & \CheckmarkBold & \XSolidBrush & \CheckmarkBold & \CheckmarkBold & \CheckmarkBold & \begin{tabular}[c]{@{}c@{}}Aerofoil,\\ MNIST\end{tabular} & 15, 500 & \begin{tabular}[c]{@{}c@{}}FedAvg, \\ HierFAVG\end{tabular} \\ \cline{2-11}
 & \multirow{4}{*}{\textbf{Both}} & \cite{yu2019distributed} & \CheckmarkBold & \XSolidBrush & \XSolidBrush & \XSolidBrush & \CheckmarkBold & \begin{tabular}[c]{@{}c@{}}CIFAR-10,\\ ATIS\end{tabular} & 16 & NULL \\ \cline{3-11} 
 &  & \cite{zhang2020federated} & \CheckmarkBold & \CheckmarkBold & \CheckmarkBold & \CheckmarkBold & \CheckmarkBold & \begin{tabular}[c]{@{}c@{}}FEMNIST, \\ Shakespeare\end{tabular} & 3500, 2288 & \begin{tabular}[c]{@{}c@{}}FedAvg, \\ C-FedAvg\end{tabular} \\ \cline{3-11} 
 &  & \cite{ye2022decentralized} & \CheckmarkBold & \XSolidBrush & \XSolidBrush & \XSolidBrush & \CheckmarkBold & CIFAR-10 & 8/16/24 & DSGD \\ \hline \toprule[0.8pt]
\multirow{26}{*}{\rotatebox{90}{\textbf{\begin{tabular}[c]{@{}c@{}} Noisy\\ fading channel\end{tabular}}}} & \multirow{15}{*}{\textbf{Down}} & \cite{zhu2019broadband} & \CheckmarkBold & \CheckmarkBold & \CheckmarkBold & \XSolidBrush & \XSolidBrush & MNIST & 200 & BAA \\ \cline{3-11} 
 &  & \cite{amiri2020federated2} & \XSolidBrush & \CheckmarkBold & \XSolidBrush & \CheckmarkBold & \XSolidBrush & MNIST & 25 & \begin{tabular}[c]{@{}c@{}}SignSGD,\\ QSGD\end{tabular} \\ \cline{3-11} 
 &  & \cite{yang2022over} & \CheckmarkBold & \CheckmarkBold & \XSolidBrush & \XSolidBrush & \XSolidBrush & MNIST & 10 & \begin{tabular}[c]{@{}c@{}}FedAvg, \\ COTAF\end{tabular} \\ \cline{3-11} 
 &  & \cite{zhang2020gradient} & \XSolidBrush & \CheckmarkBold & \XSolidBrush & \XSolidBrush & \XSolidBrush & MNIST & 10 & NULL \\ \cline{3-11} 
 &  & \cite{guo2020analog} & \CheckmarkBold & \CheckmarkBold & \XSolidBrush & \XSolidBrush & \XSolidBrush & \begin{tabular}[c]{@{}c@{}}Random \\ Gaussian matrix\end{tabular} & 5 & \begin{tabular}[c]{@{}c@{}}One-bit.\\  Open-loop\end{tabular} \\ \cline{3-11} 
 &  & \cite{sery2020analog} & \CheckmarkBold & \XSolidBrush & \XSolidBrush & \XSolidBrush & \XSolidBrush & Million Song & 100 & \begin{tabular}[c]{@{}c@{}}LAQ-WK,\\ FDM-GD,\\ ECESA-DSGD\end{tabular} \\ \cline{3-11} 
 &  & \cite{hellstrom2021over} & \CheckmarkBold & \XSolidBrush & \XSolidBrush & \XSolidBrush & \XSolidBrush & MNIST & 20 & NULL \\ \cline{3-11} 
 &  & \cite{lin2022relay} & \XSolidBrush & \CheckmarkBold & \XSolidBrush & \XSolidBrush & \XSolidBrush & FMNIST & \begin{tabular}[c]{@{}c@{}}from 2 \\ to 40\end{tabular} & \begin{tabular}[c]{@{}c@{}}FedAvg,\\ Error-free \\ channel\end{tabular} \\ \cline{2-11} 
 & \multirow{2}{*}{\textbf{Up}} & \cite{xia2021fast} & \CheckmarkBold & \XSolidBrush & \CheckmarkBold & \XSolidBrush & \CheckmarkBold & \begin{tabular}[c]{@{}c@{}}Random \\ Gaussian matrix\end{tabular} & 100 & GBMA \\ \cline{3-11} 
 &  & \cite{amiri2021convergence} & \CheckmarkBold & \CheckmarkBold & \XSolidBrush & \CheckmarkBold & \XSolidBrush & MNIST & 40 & NULL \\ \cline{2-11} 
 & \multirow{7}{*}{\textbf{Both}} & \cite{mashhadi2021fedrec} & \XSolidBrush & \CheckmarkBold & \XSolidBrush & \XSolidBrush & \XSolidBrush & \begin{tabular}[c]{@{}c@{}}Random \\ Gaussian matrix\end{tabular} & 5 & \begin{tabular}[c]{@{}c@{}}MAP bound, \\ Model-based\\ detection,   \\ CL, \\ NL\end{tabular} \\ \cline{3-11} 
 &  & \cite{wei2021federated} & \CheckmarkBold & \CheckmarkBold & \XSolidBrush & \XSolidBrush & \CheckmarkBold & \begin{tabular}[c]{@{}c@{}}MNIST, \\ CIFAR-10, \\ Shakespeare\end{tabular} & \begin{tabular}[c]{@{}c@{}}2000 10\%, \\ 100 10 \%,\\ 300, 10\%\end{tabular} & \begin{tabular}[c]{@{}c@{}}Noise-free \\ channel, \\ Equal power \\ allocation\end{tabular} \\ \cline{3-11} 
 &  & \cite{amiri2021federated} & \XSolidBrush & \XSolidBrush & \CheckmarkBold & \XSolidBrush & \CheckmarkBold & MNIST & 100 & LFL \\ \cline{3-11} 
 &  & \cite{li2021delay} & \CheckmarkBold & \CheckmarkBold & \XSolidBrush & \CheckmarkBold & \XSolidBrush & MNIST & 30 & NULL \\ \hline \toprule[0.8pt]
\end{tabular}%
}
\end{table}

\subsection{Unreliable networks}

In FL, mobile devices may frequently \textit{drop offline} due to {{unreliable network factors such as}} unstable network links, insufficient device power supply, and slow device training, resulting in a decrease in the performance and convergence rate of the entire FL system. As such, \citet{yu2019distributed} consider the unreliable network situation that all communication links have a certain packet loss rate to fail, and gives a theoretical analysis of this problem. However, it might be unreasonable to assume that each communication link has the same loss rate since it varies from the duration and packet size of different packets in reality. Consequently, \citet{zhang2020federated} design a new algorithm called ACFL to adaptively compress the information from the shared model based on the physical conditions of the current network, taking into account the drop rate and the total number of transmitted packets. Moreover, \citet{salehi2021federated} propose a different method to calculate the success probability of each link. Specifically, they utilize stochastic geometry tools to compute the loss rate and apply different weights depending on the scheduling policy and its transmission
success probability to each client during global aggregation for better performance. Furthermore, \citet{wu2020safa} propose that SAFA facilitates the influence of straggling clients and outdated models in heterogeneous scenarios. In particular, trendy and deprecated clients take the most recent global model as the base model for the subsequent round of training, but tolerated {{clients}} can continue processing their old local findings.

Since the network's physical condition is hard to measure, \citet{wu2020accelerating} propose a reliability-agnostic framework called HybridFL, which explores the situation that the reliability of clients is agnostic under strong privacy-preserving conditions and tackles it by adding regional slack factors and adjusting client selection regionally.

The previous work reckons the unreliable network issue at the \textbf{physical layer} level, whereas some work discusses it from the point of view of the \textbf{transport layer} level. As an example, \citet{mao2022safari, ye2022decentralized} consider the unreliable communication protocol, such as the user datagram protocol (UDP), rather than the reliable transportation layer protocol, such as the transmission control protocol (TCP) using a link reliability matrix to optimize mixed weights. Furthermore, the gossip-based averaging protocol is the most commonly used fault-tolerant large-scale framework \cite{lian2017can}. Nevertheless, the complexity of gossip would grow linearly with the number of clients, decelerating the convergence of FL and increasing the communication overhead. Thus, \citet{ryabinin2021moshpit} propose Moshpit All-Reduce which permits {{clients}} to dynamically select the group to which they belong, and each client only influences its current group after dropout.

\subsection{Noisy fading channels}
Recently, more and more research has focused on FL in practice, particularly over-the-air FL (OTA-FL), in which all devices transmit their data signals simultaneously through the MAC and perform computations through the wireless channel. OTA-FL could simultaneously utilize complete spectral and temporal resources and thus reduce the communication overhead in FL. However, it suffers from the noisy fading channel problem since it may cause serious problems, such as dropped packets or incorrect packet information.

Most related research aims to optimize over the \textbf{uplink} noisy fading multiple access channel (MAC).
\citet{zhu2019broadband} first assume the noisy fading channel following the {{iid}} Rayleigh fading and then design a fundamental aggregation algorithm called broadband analog aggregation (BAA) to utilize the wave superposition property of MAC for efficient update aggregation and resistance of fading channel. As an extension of BAA, \citet{amiri2020federated2} propose two more powerful and robust algorithms to tackle the fading channel issue. They create a digitally distributed SGD (D-DSGD) to select a single device for transmission in each iteration, whereas the D-DSGD lacks robustness. Thus, they develop compressed analog DSGD (CA-DSGD) to accelerate computation through sparsity and allocate the power alignment of each client at the {{server side}} for better efficiency and robustness. Nevertheless, the CA-DSGD's power alignment procedure could not be applied to the heterogeneous scenarios and controlled by each client. Subsequently, \citet{yang2022over} consider the convergence impact of noise on OTA-FL in {{non-iid}} and heterogeneous conditions and proposes a more flexible framework ACPC-OTA-FL to permit each client to adaptively calculate its transmit power level and a number of local update rounds in order to maximize the utilization of computation and communication resources. In addition, \citet{zhang2020gradient} adjust the transmit power of a device on the fly depending on aggregated gradient estimates that have been collected in the past for high-performance and reliable over-the-air computation (AirComp) over fading channels. Another gradient aggregation algorithm called analog gradient aggregation (AGA) \cite{guo2020analog} adaptively change the receiver's parameter based on data and channel state information (CSI) under fading channel conditions.
More directly, \citet{sery2020analog} claim that their proposed Gradient-Based Multiple Access (GBMA) framework does not need any power control or beamforming to get rid of fading effects, including Rayleigh, Rician, and Nakagami fading models. Instead, GBMA takes effect directly with noise distortion gradients.

Except for optimizing the problem of noisy fading uplink channel via the aggregation method and allocating transmit power, \citet{hellstrom2021over} use retransmissions after the failure of communication to reduce such estimation errors and increase the convergence rate. Furthermore, \citet{lin2022relay} investigate the problem of stragglers in fading channel situations and solves it by assigning relays to permit clients to upload their models to the relay server and mitigate the problem of stragglers.

Although most work considers the uplink as the key to communication bottleneck and supposes the downlink is error-free, imperfect \textbf{downlink} transmission could still tremendously affect the convergence and performance of FL. \citet{xia2021fast} develop FedSplit to solve the ill-conditioned problem over noisy fading channels and recover the aggregation of local updates calculated by the end devices through AirComp. Moreover, \citet{mashhadi2021fedrec} establish a data-driven universal symbol detection procedure under the downlink fading channel and a new neural network based on the maximum a-posteriori probability (MAP) detector. Furthermore, \citet{amiri2021convergence} tackle downlink fading broadcast downlink by organizing a digital approach to quantize the global model update and provides a theoretical convergence analysis for analog downlink transmission.

Comprehensively, some studies also simultaneously investigate both the noisy fading \textbf{downlink and uplink} channels.
For instance, \citet{wei2021federated} provide a rigorous convergence analysis toward standard FedAvg under {{non-iid}} client dataset, partial client participation, and noisy fading downlink and uplink channels conditions.
\citet{amiri2021federated} select a subset of clients with the highest channel gain over both the uplink and downlink fading channel conditions, which results in better global performance and more accurate information exchange.
However, the previous work does not consider the stochastic delay existing in time-varying fading channels.
In order to solve the problem, \citet{li2021delay} examine the delay distribution for wireless FL systems with uplink and downlink transmission using both synchronous and asynchronous downlink transmission strategies by exploiting the combination of saddle point approximation, extreme value theory (EVT), and large deviation theory (LDT).

\section{Communication Resource Allocation.}

Due to the difficulty posed by the heterogeneity of each device's resources, {{how to efficiently communicate model information in such scenarios}} has become a hot topic in recent years. In this section, we classify each related article according to the target of their objective functions because most of the research in this field utilizes the optimization method to solve the problem, and we list the comparison of these strategies in \Cref{Table3}.

\begin{table}[]
\centering
\caption{Classification and comparison of surveyed FL strategies on communication resource allocation. \textit{Constraints} implies the constraints of the objective function optimized in each paper.}
\label{Table3}
\resizebox{\textwidth}{!}{%
\begin{tabular}{cccccccccc}
\hline
\toprule[0.8pt]
\multirow{2}{*}{\textbf{}} & \multirow{2}{*}{\textbf{Ref.}} & \multicolumn{2}{c}{\textbf{Method}} & \multicolumn{3}{c}{\textbf{Communication}} & \multicolumn{3}{c}{\textbf{Evaluation}} \\ \cline{3-10} 
 &  & \textbf{\begin{tabular}[c]{@{}c@{}}Theoretical\\ guarantee\end{tabular}} & \textbf{non-iid} & \textbf{\begin{tabular}[c]{@{}c@{}}Client\\ selection\end{tabular}} & \textbf{Constraints} & \textbf{Partial node} & \textbf{Datasets} & \textbf{\begin{tabular}[c]{@{}c@{}}\# of\\ devices\end{tabular}} & \textbf{\begin{tabular}[c]{@{}c@{}}FL\\ Baselines\end{tabular}} \\ \hline \toprule[0.8pt]
\multirow{6}{*}{\rotatebox{90}{\textbf{\begin{tabular}[c]{@{}c@{}}Global\\ loss function\end{tabular}}}} & \cite{shi2020joint} & \CheckmarkBold & \CheckmarkBold & \CheckmarkBold & \begin{tabular}[c]{@{}c@{}}Time budget,\\ Client subset\end{tabular} & \CheckmarkBold & \begin{tabular}[c]{@{}c@{}}MNIST,\\ CIFAR-10\end{tabular} & 20 & \begin{tabular}[c]{@{}c@{}}Random\\ Proportional \\ fair\end{tabular} \\ \cline{2-10} 
 & \cite{yu2022energy} & \XSolidBrush & \XSolidBrush & \XSolidBrush & Energy budget & \XSolidBrush & MNIST & 50 & Loss, time \\ \cline{2-10} 
 & \cite{mahmoudi2022fedcau} & \CheckmarkBold & \CheckmarkBold & \XSolidBrush & Latency budget & \XSolidBrush & MNIST & 50, 100 & \begin{tabular}[c]{@{}c@{}}LAQ,\\ Top-k\end{tabular} \\ \cline{2-10} 
 & \cite{wang2019adaptive} & \CheckmarkBold & \XSolidBrush & \XSolidBrush & \begin{tabular}[c]{@{}c@{}}Different types\\ of resources\end{tabular} & \XSolidBrush & MNIST & \begin{tabular}[c]{@{}c@{}}from 5\\ to 500\end{tabular} & \begin{tabular}[c]{@{}c@{}}SVM,\\ K-means\end{tabular} \\ \hline \toprule[0.8pt]
\multirow{8}{*}{\rotatebox{90}{\textbf{Time consumption}}} & \cite{chen2020convergence} & \CheckmarkBold & \XSolidBrush & \CheckmarkBold & \begin{tabular}[c]{@{}c@{}}Different resources, \\ Client subset\end{tabular} & \CheckmarkBold & MNIST & \begin{tabular}[c]{@{}c@{}}from 3\\ to 20\end{tabular} & FedAvg \\ \cline{2-10} 
 & \cite{yang2020delay} & \CheckmarkBold & \XSolidBrush & \XSolidBrush & \begin{tabular}[c]{@{}c@{}}Bandwidth,\\ Time budget,\\ Uploaded data size, \\ Transmit power\end{tabular} & \XSolidBrush & \begin{tabular}[c]{@{}c@{}}Real open\\ blog feedback\\ dataset\end{tabular} & 50 & \begin{tabular}[c]{@{}c@{}}EB-FDMA,\\ FE-FDMA,\\ TDMA\end{tabular} \\ \cline{2-10} 
 & \cite{lu2020low} & \XSolidBrush & \XSolidBrush & \XSolidBrush & \begin{tabular}[c]{@{}c@{}}Digital twins,\\ Subchannel,\\ Training batch size\end{tabular} & \XSolidBrush & CIFAR-10 & 100 & FedAvg \\ \hline \toprule[0.8pt]
\multirow{6}{*}{\rotatebox{90}{\textbf{Client selection}}} & \cite{nishio2019client} & \XSolidBrush & \CheckmarkBold & \CheckmarkBold & \begin{tabular}[c]{@{}c@{}}Time budget, \\ Communication \\ rounds\end{tabular} & \CheckmarkBold & \begin{tabular}[c]{@{}c@{}}CIFAR-10,\\ FMNIST\end{tabular} & \begin{tabular}[c]{@{}c@{}}1000, \\ 10\%\end{tabular} & FedLim \\ \cline{2-10} 
 & \cite{chen2022federated} & \CheckmarkBold & \CheckmarkBold & \CheckmarkBold & \begin{tabular}[c]{@{}c@{}}Power budget, \\ Bandwidth, \\ Time budget\end{tabular} & \CheckmarkBold & MNIST & 10 & FedAvg \\ \cline{2-10} 
 & \cite{zhao2021system} & \XSolidBrush & \XSolidBrush & \XSolidBrush & \begin{tabular}[c]{@{}c@{}}Bandwidth,\\ Time budget\end{tabular} & \XSolidBrush & \begin{tabular}[c]{@{}c@{}}MNIST,\\ FMNIST\end{tabular} & 30 & BBA \\ \hline \toprule[0.8pt]
\multirow{2}{*}{\rotatebox{90}{\textbf{\begin{tabular}[c]{@{}c@{}}Energy\\ consumption\end{tabular}}}} & \cite{zeng2020energy} & \XSolidBrush & \XSolidBrush & \CheckmarkBold & \begin{tabular}[c]{@{}c@{}}Bandwidth,\\ Time budget\end{tabular} & \CheckmarkBold & MNIST & 50 & \begin{tabular}[c]{@{}c@{}}Optimal/\\ uniform\\ bandwidth\\ allocation\end{tabular} \\ \cline{2-10} 
 & \cite{yang2020energy} & \XSolidBrush & \CheckmarkBold & \CheckmarkBold & \begin{tabular}[c]{@{}c@{}}Bandwidth,\\ Time budget\end{tabular} & \CheckmarkBold & \begin{tabular}[c]{@{}c@{}}Real open\\ blog feedback\\ dataset\end{tabular} & 50 & TDMA \\ \hline \toprule[0.8pt]
\multirow{3}{*}{\rotatebox{90}{\textbf{\begin{tabular}[c]{@{}c@{}}Hybrid\\ objective\end{tabular}}}} & \cite{imteaj2020fedar} & \XSolidBrush & \XSolidBrush & \CheckmarkBold & \begin{tabular}[c]{@{}c@{}}Resource \\ availability, \\ Trust, \\ Client subset\end{tabular} & \CheckmarkBold & MNIST & 12 & NULL \\ \cline{2-10} 
 & \cite{zaw2021energy} & \XSolidBrush & \XSolidBrush & \XSolidBrush & \begin{tabular}[c]{@{}c@{}}Time budget, \\ Energy budget\end{tabular} & \XSolidBrush & MNIST & 50 & \begin{tabular}[c]{@{}c@{}}Loss, \\ energy\end{tabular} \\ \hline \toprule[0.8pt]
\end{tabular}%
}
\end{table}

\textbf{Global Loss Function.} \citet{shi2020joint} establish an associated client scheduling and bandwidth allocation strategy based on optimizing the trade-off between latency and the number of training rounds. Additionally, \citet{yu2022energy} consider the hybrid vertically and horizontally partitioned client dataset and optimize both the communication and computational energy consumption to develop a scheduling {{optimization problem}} for a participant {{jointly}} and then minimize the global loss function. Furthermore, \citet{mahmoudi2022fedcau} propose an iteration-termination {{criterion}} FedCau by optimizing the computation and communication latency. Furthermore, they also combine FedCau with Top-q and LAQ compression methods to further reduce communication overhead.
 
However, the above papers may lack some generality since they all specify the type of resources, such as bandwidth and computational energy. To this end, \citet{wang2019adaptive} propose a more general method to dynamically minimize the loss function under a resource budget constraint without specifying a resource type and analyze the convergence bound for federated learning with a non-iid dataset. 
    
\textbf{Time Consumption.} \citet{chen2020convergence} simultaneously minimize the communication time and convergence time to construct a user selection and power allocation framework. However, it only considers the transmission time constraint. Subsequently, \citet{yang2020delay} take the computation time into account and also develop a bisection search algorithm to find the optimal solution. Furthermore, \citet{lu2020low} also consider the cost of the block validation process and introduce the digital twin wireless networks (DTWN) to alleviate the resource-constrained FL task in real-world environments.
    
\textbf{Client Selection.} A greedy scheduling strategy is proposed by \citet{nishio2019client}, which schedules as many devices as possible within a given time frame in each round. However, the deadlines are chosen experimentally, and it is tough to adapt to dynamic channels and device computing power. 
To this end, \citet{chen2022federated} propose CEFL to reuse outdated local model parameters for client scheduling and theoretically analyze its convergence and communication characteristics. Nonetheless, the previous articles do not consider channel state information (CSI). Thus, \citet{zhao2021system} propose two bandwidth allocation methods based on CSI and particle swarm optimization (PSO) and transform the optimization problem from minimizing global loss into maximizing the number of active clients.
    
\textbf{Energy Consumption.} \citet{zeng2020energy} propose an energy-efficient joint bandwidth allocation and scheduling strategy by minimizing the total energy consumption. However, they assume that all clients have the same {{model structure}} and thus ignore the computational energy. Nevertheless, \citet{yang2020energy} further consider the computational energy and derive closed-form solutions for computation and transmission resources of a low-complexity iteration algorithm.
    
\textbf{Hybrid Objective.} {{In the meanwhile, some research focuses on optimizing multiple objects in the objective function.}} For example, \citet{imteaj2020fedar} try to optimize the client resources such as memory, bandwidth, and battery life under the malicious client and straggler issue, but no theoretical guarantees are provided. Comprehensively, \citet{zaw2021energy} aim to optimize the global loss function and the entire communication and computation time and then reformulate the problem as a Generalized Nash Equilibrium Problem (GNEP) to establish a comprehensive theoretical convergence analysis.

\section{Conclusions and Future Directions}

Due to the complex wireless network environment, the heterogeneity of device data, the difference of device {{capabilities}}, and other factors, how to efficiently perform FL training in the edge network is a key issue currently facing. In this review, we discuss the solutions to this issue in three aspects: communication efficiency, communication environment, and communication resource allocation.

First, we present the communication efficiency strategies in FL. Specifically, we dive into the three main communication-efficient FL approaches: quantization, sparsification, and knowledge distillation. For each strategy, we examine its benefits, drawbacks, and prospective logical development trend in order to make the entire paragraph logically coherent.

Second, we discuss the influence of the harsh FL communication environment. We divide the harsh environment into two parts: 1) unreliable network: the connections between clients and servers are not guaranteed to succeed and thus may cause clients to go offline; 2) noisy fading channels: the communication channels have noise and fade along with the transmission distance.

Finally, we present the communication resource allocation algorithms, which mainly target solving an optimization problem based on some realistic constraints. Therefore, we classify such algorithms according to the optimization objective of the optimization problem. In particular, we find that the global loss function, time consumption, energy consumption, client selection, and hybrid objectives are the most frequently used objectives.

{{Even if the existing research is continually being refined, there are still some future directions we can investigate in this subject.}}

\textbf{Computation Consumption.} As mentioned earlier, many communication compression algorithms run on the client side. Although there are some studies discussing computing time and computing power consumption in Section 5, few studies work on the specific computing resource consumption in the compression algorithm, which is exactly significant for resource-limited mobile devices. {{One work \citet{mishchenko2022intsgd} is related to this field, in which they propose a quantized parallel SGD where the gradient coordinates are rounded after scaling. Further work can be performed to extend it to popular optimization methods such as Adam.}}

\textbf{A Comprehensive Communication System.} Most of the previous studies considered only one compression method, and some studies considered the combination of two compression methods. However, it is possible that this combination is not the most efficient. In future work, the integration of multiple model compression methods to formulate an overall FL compression system and finding the balance of each compression method theoretically may become the next outlet for communication-efficient FL.

  \textbf{Exploring Privacy while Compressing.} Privacy is also a widely studied field in FL. {{As mentioned in the introduction section, FL without privacy-preserving procedures may nevertheless leak local private information through the transmission of model parameters or model gradients.}} In order to solve this kind of data privacy leakage problem, some existing research uses cryptography to encrypt the model parameters sent by each participant to defend against model inference attacks, such as LDP and SMPC introduced in Section 3.3. In addition, other protective strategies, such as VerifyNet \cite{xu2019verifynet} and adversarial training \cite{tramer2017ensemble}, are effective during the training stage and preserve {{clients'}} privacy. {{Nonetheless, a substantial quantity of local calculation overhead will make the entire FL process extremely sluggish. Due to the heterogeneity of local client computing resources, some clients may take a long time to perform encryption algorithms while other clients can only wait, thereby further causing communication congestion. Consequently, the demand for efficient privacy-preserving strategies is increasing. }}Combining compression techniques and privacy-preserving techniques is one way to make the entire FL framework {{both efficient and secure}}. As an illustration, federated distillation only needs to transmit a few fully-connected layers or even logits between clients and the server, which simultaneously compresses the updated information and protects the data privacy.

\textbf{Asynchronous Aggregation.} {{In this review paper, we mostly concentrate on synchronous FL aggregation. However, due to the fact that the available computation, communication, and data volumes on many work nodes are sometimes distinct, the time that work nodes transmit model parameters of local training at each round may vary from one to another}}. This will cause the parameter server to prolong the training time due to waiting for the slow node to upload the parameters (i.e., the straggler problem). In addition, since the local data on multiple work nodes usually may not follow the same probability distribution and their local models may also be different, it will cause the local models of different work nodes to converge in different directions, reducing the overall training speed. Therefore, the asynchronous aggregate update algorithm should also be considered when considering the communication overhead to make the model converge faster and obtain better model performance.

\textbf{Integration with 5G and Beyond.} The rapid advances in wireless communication technologies also enhance the development of FL technologies.  The bandwidth and processing speed constraints of FL are expected to be lifted by the expansion of the 5G network across the globe \cite{abd2019efficient}.  5G network is estimated to provide a 10-100x increase in bandwidth and a 5-10x decrease in latency, enabling more devices on the Internet of Things (IoT) and the Internet of Vehicles (IoV) devices to learn from each other with federated learning. For example, China Mobile proposed the integration of FL with the distributed intelligent network architecture \cite{White_paper}. Internationally, industrial standards on FL in 5G networks have also been proposed in \cite{3GPP}. 6G communications can achieve ultra-low delay, superior reliability, and high energy efficiency \cite{xiao2020towards}. Since 5G and 6G networks will enable a massive number of heterogeneous devices of higher intelligence learned from explosive data, the optimization of communication efficiency over the entire FL network will undoubtedly be more challenging.  Heterogeneity in device capabilities, network connectivity, and energy level will also lead to complex network behaviors and cause more vulnerabilities and failures \cite{liu2020federated}. Therefore, it is important to perform real-life simulations and efficient training to accelerate convergence and identify bottlenecks in key applications and environments, such as AR/VR and intelligent transportation \cite{liu2020privacy}, to address these issues.

\bibliography{mybibfile}


\end{document}